\documentclass{article}%
\usepackage{amssymb}
\usepackage{graphicx}%
\usepackage{amsmath}%
\usepackage{amsfonts}

\begin{document}

\begin{center}
{\LARGE Correlation Functions of Complex Matrix Models}

\bigskip

\bigskip

{\LARGE M.C. Berg\`{e}re}

Service de Physique th\'{e}orique, CEA-Saclay

F-91191 Gif sur Yvette, France

email: bergere@spht.saclay.cea.fr

\bigskip

October 29, 2005

\bigskip

\bigskip

\textbf{Abstract}

\bigskip
\end{center}

\qquad For a restricted class of potentials (harmonic+Gaussian potentials), we
express the resolvent integral for the correlation functions of simple traces
of powers of complex matrices of size $N$, in term of a determinant; this
determinant is function of four kernels constructed from the orthogonal
polynomials corresponding to the potential and from their Cauchy transform.
The correlation functions are a sum of expressions attached to a set of fully
packed oriented loops configurations; for rotational invariant systems,
explicit expressions can be written for each configuration and more
specifically for the Gaussian potential, we obtain the large $N$\ expansion
('t\ Hooft expansion)\ and the so-called BMN limit.

\bigskip

\bigskip

\section{{\protect\LARGE Introduction}}

\bigskip

\qquad Given a potential $V\left(  M,M^{+}\right)  \ $where $M$ is a $N\times
N\ $matrix with complex elements, we define the partition function as%
\begin{equation}
Z=\int dM\ dM^{+}\ e^{-Tr\ V\left(  M,M^{+}\right)  \ }\tag{1.1}%
\end{equation}
where we integrate over all matrix elements. We define the correlation
functions as%
\begin{equation}
<%
{\displaystyle\prod\limits_{i=1}^{n}}
O_{i}\left(  M,M^{+}\right)  >=\frac{1}{Z}\int dM\ dM^{+}\
{\displaystyle\prod\limits_{i=1}^{n}}
O_{i}\left(  M,M^{+}\right)  \ \ e^{-Tr\ V\left(  M,M^{+}\right)  \ }\tag{1.2}%
\end{equation}
Most observables $O\left(  M,M^{+}\right)  \ $can be expressed in terms of
traces of product of the matrices $M$ and $M^{+}$; among these traces, we
distinguish between "mixed traces" like $Tr\ \left[  M^{J_{1}}\left(
M^{+}\right)  ^{K_{1}}\ M^{J_{2}}\ \left(  M^{+}\right)  ^{K_{2}}...\right]  $
which are also called "words", and simple traces where there is only one type
of matrix ($M$ or $M^{+}$) inside the trace. In this publication, we restrict
ourselves to the case of simple traces%
\begin{equation}
<%
{\displaystyle\prod\limits_{i=1}^{p}}
\left[  Tr\ M^{J_{i}}\right]  \ \
{\displaystyle\prod\limits_{i=1}^{q}}
\left[  Tr\ \left(  M^{+}\right)  ^{K_{i}}\right]
>\ \ \ \ \ \ \ \ \ \ \ \ J_{i}\ \text{and }K_{i}>0\tag{1.3}%
\end{equation}
for a restricted class of potentials (described below). We now explain the
method used to obtain the above correlation functions.

\bigskip

If certain conditions on the potential $V\left(  M,M^{+}\right)  $ are
satisfied, we were able in Ref.$\left[  1\right]  $ to calculate the following
expectation values%
\begin{equation}
\frac{1}{Z}\int dM\ dM^{+}\ \frac{%
{\displaystyle\prod\limits_{i=1}^{L_{2}}}
\det\left(  \eta_{i}^{\ast}-M^{+}\right)  }{%
{\displaystyle\prod\limits_{i=1}^{M_{2}}}
\det\left(  x_{i}^{\ast}-M^{+}\right)  }\ \ \frac{%
{\displaystyle\prod\limits_{i=1}^{L_{1}}}
\det\left(  \xi_{i}-M\right)  }{%
{\displaystyle\prod\limits_{i=1}^{M_{1}}}
\det\left(  y_{i}-M\right)  }\ \ e^{-Tr\ V\left(  M,M^{+}\right)  \ }\tag{1.4}%
\end{equation}
for any numbers $L_{1},\ M_{1},\ L_{2},\ M_{2}\ $of determinants. The general
result is essentially a determinant constructed from four kernels, each kernel
is a function of the external sources $\xi_{i},\ \eta_{i}^{\ast}%
,\ y_{i},\ x_{i}^{\ast}$ which enter the orthogonal polynomials relative to
the potential $V\left(  z,z^{\ast}\right)  \ $and their Cauchy transform.\ The
structure of this determinant is very dependant of the relative values of
$L_{1},\ M_{1},\ L_{2},\ M_{2}$. In this publication we are only interested in
the special case where $L_{1}=M_{1},\ L_{2}=M_{2}$\ since in that case, if we
perform the derivatives $%
{\displaystyle\prod\limits_{i=1}^{M_{2}}}
\left(  \frac{\partial}{\partial\eta_{i}^{\ast}}\right)  _{x_{i}^{\ast}%
=\eta_{i}^{\ast}}\ \
{\displaystyle\prod\limits_{i=1}^{M_{1}}}
\left(  \frac{\partial}{\partial\xi_{i}}\right)  _{y_{i}=\xi_{i}}\ $on (1.4)
we obtain%
\begin{equation}
\frac{1}{Z}\int dM\ dM^{+}\
{\displaystyle\prod\limits_{i=1}^{M_{2}}}
Tr\frac{1}{\eta_{i}^{\ast}-M^{+}}\
{\displaystyle\prod\limits_{i=1}^{M_{1}}}
Tr\frac{1}{\xi_{i}-M}\ \ e^{-Tr\ V\left(  M,M^{+}\right)  \ }\tag{1.5}%
\end{equation}
The above resolvent integrals are the generating functionals for the
integrals
\begin{equation}
\frac{1}{Z}\int dM\ dM^{+}\
{\displaystyle\prod\limits_{i=1}^{M_{2}}}
Tr\ \left[  \left(  M^{+}\right)  ^{K_{i}}\right]  \ \
{\displaystyle\prod\limits_{i=1}^{M_{1}}}
Tr\ \left[  \left(  M\right)  ^{J_{i}}\right]  \ \ \ e^{-Tr\ V\left(
M,M^{+}\right)  \ }\tag{1.6}%
\end{equation}
which describe the correlation functions (1.3).

\bigskip

Let us now describe the conditions required by the potential $V\left(
M,M^{+}\right)  .$\ Any matrix $M$ can be triangularized by a unitary matrix%
\begin{equation}
M=U\ \left[  D+T\right]  \ U^{+}\tag{1.7}%
\end{equation}
where $D$ is the diagonal matrix of the eigenvalues $\left(  z_{1}%
,...,z_{N}\right)  \ $of $M,\ $and $T$ is an upper strictly triangular matrix.
Consequently, any power of $M$ and $M^{+}\ $is of the form%
\begin{align}
M^{k}  & =U\ \left[  D^{k}+T_{k}\right]  \ U^{+}\tag{1.8.a}\\
\left(  M^{+}\right)  ^{k}  & =U\ \left[  \left(  D^{+}\right)  ^{k}+T_{k}%
^{+}\right]  \ U^{+}\tag{1.8.b}%
\end{align}
where $T_{k}$ is an upper triangular matrix which also depends of the
eigenvalues $z_{i}\ $for $k>1$. Clearly, we have%
\begin{align}
Tr\ M^{k}  & =\sum_{i=1}^{N}z_{i}^{k}\tag{1.9.a}\\
Tr\ \left(  M^{+}\right)  ^{k}  & =\sum_{i=1}^{N}\left(  z_{i}^{\ast}\right)
^{k}\tag{1.9.b}%
\end{align}
On the other hand,
\begin{equation}
Tr\ MM^{+}=\sum_{i=1}^{N}z_{i}\ z_{i}^{\ast}+Tr\ TT^{+}\tag{1.10}%
\end{equation}
In the transformation (1.7),%

\begin{align}
\det\left(  \xi_{i}-M\right)   & =\det\left(  \xi_{i}-D\right) \tag{1.11.a}\\
\det\left(  \eta_{i}^{\ast}-M^{+}\right)   & =\det\left(  \eta_{i}^{\ast
}-D^{+}\right) \tag{1.11.b}%
\end{align}
and the integration measure becomes%
\begin{equation}
dM\ dM^{+}=dU\ dT\ dT^{+}\ dD\ dD^{+}\
{\displaystyle\prod\limits_{i<j}}
\left\vert z_{i}-z_{j}\right\vert ^{2}\tag{1.12}%
\end{equation}
We note that $dM\ dM^{+}$ represents $2N^{2}$\ variables of integration while
$dU\ $represents $N\left(  N-1\right)  $ variables of integration$,\ dT$
$dT^{+}$ represents $N\left(  N-1\right)  $\ variables of integration\ and
$dD\ dD^{+}$ represents $2N$\ variables of integration.\ The Jacobian of the
transformation is given by the product of two Vandermonde determinants
$\Delta\left(  z\right)  \ \Delta\left(  z^{\ast}\right)  \ $where%
\begin{equation}
\Delta\left(  z\right)  =%
{\displaystyle\prod\limits_{i<j}}
\left(  z_{i}-z_{j}\right)  =\left\vert
\begin{array}
[c]{ccc}%
\pi_{N-1}\left(  z_{1}\right)  & ... & \pi_{N-1}\left(  z_{N}\right) \\
... & ... & ...\\
\pi_{0}\left(  z_{1}\right)  & ... & \pi_{0}\left(  z_{N}\right)
\end{array}
\right\vert \tag{1.13}%
\end{equation}
and where the polynomials $\left\{  \pi_{0}\left(  z\right)  ,\ \pi_{1}\left(
z\right)  ,\ \pi_{2}\left(  z\right)  ,...,\pi_{N-1}\left(  z\right)
\ \right\}  $are any set of monic polynomials\ of successive degree
$0,1,2,...,N-1.$

\bigskip

Consequently, if we choose the potential $V\left(  M,M^{+}\right)  $ to be of
the form%
\begin{equation}
V\left(  M,M^{+}\right)  =MM^{+}+V\left(  M\right)  +\overline{V}\left(
M^{+}\right) \tag{1.14}%
\end{equation}
we may separate the integrations over the unitary group $dU$ and over the
upper triangular part $dT$ $dT^{+}\ e^{-Tr\ TT^{+}}$\ and these two
contributions cancel in the ratio of the numerator and the denominator of
(1.6). We are left with integrals over the eigenvalues only.

\bigskip

More generally, we consider the integrals$\ Z_{N},\ I_{N}$ and$\ J_{N}\ $and a
potential$\ $

$V\left(  z,z^{\ast}\right)  $ such that the following integrals exist
\begin{align}
Z_{N}  & =\int d\mu\left(  z,z^{\ast}\right)  \ \ \ \tag{1.15.a}\\
I_{N}  & =\frac{1}{Z_{N}}\int d\mu\left(  z,z^{\ast}\right)  \ \
{\displaystyle\prod\limits_{j=1}^{N}}
\left\{  \frac{%
{\displaystyle\prod\limits_{i=1}^{L_{2}}}
\left(  z_{j}^{\ast}-\eta_{i}^{\ast}\right)  }{%
{\displaystyle\prod\limits_{i=1}^{M_{2}}}
\left(  z_{j}^{\ast}-x_{i}^{\ast}\right)  }\ \ \frac{%
{\displaystyle\prod\limits_{i=1}^{L_{1}}}
\left(  z_{j}-\xi_{i}\right)  }{%
{\displaystyle\prod\limits_{i=1}^{M_{1}}}
\left(  z_{j}-y_{i}\right)  }\right\} \nonumber\\
& \tag{1.15.b}\\
J_{N}  & =\frac{1}{Z_{N}}\int d\mu\left(  z,z^{\ast}\right)
{\displaystyle\prod\limits_{i=1}^{M_{2}}}
\left\{  \sum_{j=1}^{N}\frac{1}{z_{j}^{\ast}-\eta_{i}^{\ast}}\right\}
{\displaystyle\prod\limits_{i=1}^{M_{1}}}
\left\{  \sum_{j=1}^{N}\frac{1}{z_{i}-\xi_{i}}\right\}  \ \nonumber\\
& \tag{1.15.c}\\
d\mu\left(  z,z^{\ast}\right)   & =%
{\displaystyle\prod\limits_{i=1}^{N}}
dz_{i}\ dz_{i}^{\ast}\ \
{\displaystyle\prod\limits_{i<j}}
\left\vert z_{i}-z_{j}\right\vert ^{2}\ e^{-\sum_{i=1}^{N}V\left(  z_{i}%
,z_{i}^{\ast}\right)  \ }\tag{1.15.d}%
\end{align}
The integrals $J_{N}$ are generating functionals for the integrals
\begin{equation}
\frac{1}{Z_{N}}\int d\mu\left(  z,z^{\ast}\right)  \ \
{\displaystyle\prod\limits_{i=1}^{M_{2}}}
\left\{  \sum_{j=1}^{N}\left(  z_{j}^{\ast}\right)  ^{K_{i}}\right\}  \
{\displaystyle\prod\limits_{i=1}^{M_{1}}}
\left\{  \sum_{j=1}^{N}\left(  z_{j}\right)  ^{J_{i}}\right\}  \ \ \tag{1.16}%
\end{equation}

\bigskip

\bigskip

A second condition on $V\left(  M,M^{+}\right)  $ is that $V\left(  z,z^{\ast
}\right)  $ is real and admits a set of orthogonal polynomials so that the
technique of Ref$\left[  1\right]  $ can be applied (the reality of $V\left(
z,z^{\ast}\right)  $ could be forgotten at the price of introducing
biorthogonal polynomials but we shall not consider this case). We introduce
the infinite set $\left\{  p_{n}\left(  z\right)  \right\}  $ of orthogonal,
monic polynomials such that%
\begin{equation}
\int d^{2}z\ \ p_{m}^{\ast}\left(  z\right)  \ \ p_{n}\left(  z\right)
\ \ e^{-V\left(  z,z^{\ast}\right)  }=h_{n}\ \delta_{nm}\tag{1.17}%
\end{equation}
where $\ p_{m}^{\ast}\left(  z\right)  $ is a short notation for $\left[
p_{m}\left(  z\right)  \right]  ^{\ast}.\ $Then, we consider the measure
\begin{equation}
d\mu\left(  z,z^{\ast};\xi_{i},\eta_{i}^{\ast};y_{i},x_{i}^{\ast}\right)
=d^{2}z\ \frac{%
{\displaystyle\prod\limits_{i=1}^{L_{2}}}
\left(  z^{\ast}-\eta_{i}^{\ast}\right)  }{%
{\displaystyle\prod\limits_{i=1}^{M_{2}}}
\left(  z^{\ast}-x_{i}^{\ast}\right)  }\ \ \frac{%
{\displaystyle\prod\limits_{i=1}^{L_{1}}}
\left(  z-\xi_{i}\right)  }{%
{\displaystyle\prod\limits_{i=1}^{M_{1}}}
\left(  z-y_{i}\right)  }\ \ e^{-V\left(  z,z^{\ast}\right)  }\tag{1.18}%
\end{equation}
We proved in Ref.$\left[  1\right]  \ $the existence and we constructed
explicitely the set of monic, biorthogonal polynomials $q_{n}\left(  z;\xi
_{i},\eta_{i}^{\ast};y_{i},x_{i}^{\ast}\right)  $ and $q_{n}^{\ast}\left(
z;\eta_{i},\xi_{i}^{\ast};x_{i},y_{i}^{\ast}\right)  $ satisfying%
\begin{equation}
\int d\mu\left(  z,z^{\ast};\xi_{i},\eta_{i}^{\ast};y_{i},x_{i}^{\ast}\right)
\ q_{m}^{\ast}\left(  z;\eta_{i},\xi_{i}^{\ast};x_{i},y_{i}^{\ast}\right)
\ \ q_{n}\left(  z;\xi_{i},\eta_{i}^{\ast};y_{i},x_{i}^{\ast}\right)
=\left\Vert q_{n}\right\Vert ^{2}\ \ \delta_{nm}\tag{1.19}%
\end{equation}

This construction is a generalization of Christoffel's result $\left[
2\right]  $ which shows that given a positive Borel measure of one variable
$d\mu\left(  x\right)  $ on the real line and its infinite set of orthogonal
polynomials, it is possible to construct an infinite set of orthogonal
polynomials for the measure%
\begin{equation}
d\mu\left(  x;\xi_{i}\right)  =d\mu\left(  x\right)  \
{\displaystyle\prod\limits_{i=1}^{L}}
\left(  x-\xi_{i}\right) \tag{1.20}%
\end{equation}
This result was extended to measures with external sources at the denominator
by Uvarov $\left[  3\right]  $ in 1969 and recently by Fyodorov and Strahov
$\left[  4\right]  .$ In 2003, Akemann and Vernizzi extended Christoffel's
result to measures on the complex plane with external sources at the numerator
$\left[  5\right]  $ (see also Ref.$\left[  6\right]  $); their work was
generalized to sources at the denominator as well in Ref.$\left[  1\right]
\ $and $\left[  7\right]  $.

The consequence of the existence of biorthogonal polynomials for the measure
(1.18) is that%
\begin{align}
Z_{N}  & =N!\
{\displaystyle\prod\limits_{i=0}^{N-1}}
\ h_{i}\tag{1.21.a}\\
Z_{N}\ \ I_{N}  & =N!\
{\displaystyle\prod\limits_{i=0}^{N-1}}
\ \left\Vert q_{i}\right\Vert ^{2}\tag{1.21.b}%
\end{align}

In the case $L_{1}=M_{1},\ L_{2}=M_{2}$, the pseudonorms $\left\Vert
q_{i}\right\Vert ^{2}$ are found in Ref.$\left[  1\right]  \ $to be the ratio
of two determinants%
\begin{equation}
\left\Vert q_{i}\right\Vert ^{2}=h_{i}\ \frac{D_{i}}{D_{i-1}}\ \ \ \ \ i\geq
0\tag{1.22}%
\end{equation}
where\bigskip$\ D_{n}$ is the determinant
\begin{align}
& \left\vert
\begin{array}
[c]{cccccc}%
N_{n}\left(  \xi_{1},y_{1}\right)  & ... & N_{n}\left(  \xi_{M_{1}}%
,y_{1}\right)  & A_{n}\left(  x_{1}^{\ast},y_{1}\right)  & ... & A_{n}\left(
x_{M_{2}}^{\ast},y_{1}\right) \\
... & ... & ... & ... & ... & ...\\
N_{n}\left(  \xi_{1},y_{M_{1}}\right)  & ... & N_{n}\left(  \xi_{M_{1}%
},y_{M_{1}}\right)  & A_{n}\left(  x_{1}^{\ast},y_{M_{1}}\right)  & ... &
A_{n}\left(  x_{M_{2}}^{\ast},y_{M_{1}}\right) \\
K_{n}\left(  \xi_{1},\eta_{1}^{\ast}\right)  & ... & K_{n}\left(  \xi_{M_{1}%
},\eta_{1}^{\ast}\right)  & N_{n}^{\ast}\left(  \eta_{1},x_{1}\right)  & ... &
N_{n}^{\ast}\left(  \eta_{1},x_{M_{2}}\right) \\
... & ... & ... & ... & ... & ...\\
K_{n}\left(  \xi_{1},\eta_{M_{2}}^{\ast}\right)  & ... & K_{n}\left(
\xi_{M_{1}},\eta_{M_{2}}^{\ast}\right)  & N_{n}^{\ast}\left(  \eta_{M_{2}%
},x_{1}\right)  & ... & N_{n}^{\ast}\left(  \eta_{M_{2}},x_{M_{2}}\right)
\end{array}
\right\vert \nonumber\\
& \tag{1.23}%
\end{align}
and $D_{-1}\ $is defined in (1.32). Consequently, in the case $L_{1}%
=M_{1},\ L_{2}=M_{2}$ \ \ the integral $I_{N}\ $is found to be
\begin{equation}
I_{N}=\left(  -\right)  ^{\frac{M_{1}\left(  M_{1}-1\right)  }{2}}\ \left(
-\right)  ^{\frac{M_{2}\left(  M_{2}-1\right)  }{2}}\ \ \ \frac{\
{\displaystyle\prod\limits_{i,j=1}^{M_{2}}}
\left(  x_{i}^{\ast}-\eta_{j}^{\ast}\right)  \
{\displaystyle\prod\limits_{i,j=1}^{M_{1}}}
\left(  y_{i}-\xi_{j}\right)  }{\Delta\left(  x^{\ast}\right)  \ \Delta\left(
y\right)  \ \Delta\left(  \eta^{\ast}\right)  \ \Delta\left(  \xi\right)
\ }\ \ \ D_{N-1}\tag{1.24}%
\end{equation}
Then, by application of the derivatives $%
{\displaystyle\prod\limits_{i=1}^{M_{2}}}
\left(  \frac{\partial}{\partial\eta_{i}^{\ast}}\right)  _{x_{i}^{\ast}%
=\eta_{i}^{\ast}}\ \
{\displaystyle\prod\limits_{i=1}^{M_{1}}}
\left(  \frac{\partial}{\partial\xi_{i}}\right)  _{y_{i}=\xi_{i}}\ $on
\ (1.24), we proved in Ref.$\left[  1\right]  $ that%
\begin{equation}
J_{N}=\ "D_{N-1}"\tag{1.25}%
\end{equation}
where$\ "D_{n}"$ is the "subtracted determinant"%
\begin{equation}
\ \left\vert
\begin{array}
[c]{cccccc}%
H_{n}\left(  \xi_{1},\xi_{1}\right)  & ... & N_{n}\left(  \xi_{M_{1}},\xi
_{1}\right)  & A_{n}\left(  \eta_{1}^{\ast},\xi_{1}\right)  & ... &
A_{n}\left(  \eta_{M_{2}}^{\ast},\xi_{1}\right) \\
... & ... & ... & ... & ... & ...\\
N_{n}\left(  \xi_{1},\xi_{M_{1}}\right)  & ... & H_{n}\left(  \xi_{M_{1}}%
,\xi_{M_{1}}\right)  & A_{n}\left(  \eta_{1}^{\ast},\xi_{M_{1}}\right)  &
... & A_{n}\left(  \eta_{M_{2}}^{\ast},\xi_{M_{1}}\right) \\
K_{n}\left(  \xi_{1},\eta_{1}^{\ast}\right)  & ... & K_{n}\left(  \xi_{M_{1}%
},\eta_{1}^{\ast}\right)  & H_{n}^{\ast}\left(  \eta_{1},\eta_{1}\right)  &
... & N_{n}^{\ast}\left(  \eta_{M_{2}},\eta_{1}\right) \\
... & ... & ... & ... & ... & ...\\
K_{n}\left(  \xi_{1},\eta_{M_{2}}^{\ast}\right)  & ... & K_{n}\left(
\xi_{M_{1}},\eta_{M_{2}}^{\ast}\right)  & N_{n}^{\ast}\left(  \eta_{1}%
,\eta_{M_{2}}\right)  & ... & H_{n}^{\ast}\left(  \eta_{M_{2}},\eta_{M_{2}%
}\right)
\end{array}
\right\vert ^{"}\tag{1.26}%
\end{equation}

\bigskip

The determinant $D_{n}\ $is expressed in terms of four kernels $K_{n}%
,\ N_{n},N_{n}^{\ast},\ A_{n}$ which are defined with some of their properties
at the end of this introduction. The determinant $"D_{n}"$ is obtained from
the determinant $D_{n}$ by changing all kernels $N_{n}$\ on\ the diagonal into
$H_{n},$ changing all $y_{k}$ into $\xi_{k},$ all $x_{k}^{\ast}$ into
$\eta_{k}^{\ast}\ $and finally by ignoring all double poles at $\xi_{j}%
=\xi_{k}$ and at $\eta_{j}^{\ast}=\eta_{k}^{\ast}$ (this operation is denoted
by "\ \ ") as we develop the determinant $\left(  1.26\right)  \ $with terms
like $N_{n}\left(  \xi_{j},\xi_{k}\right)  \ N_{n}\left(  \xi_{k},\xi
_{j}\right)  $ or $N_{n}^{\ast}\left(  \eta_{k},\eta_{j}\right)  \ N_{n}%
^{\ast}\left(  \eta_{j},\eta_{k}\right)  $. In fact, from (1.30.b) we observe
that there is no single poles either at $\xi_{j}=\xi_{k}$ or at $\eta
_{j}^{\ast}=\eta_{k}^{\ast}$ since the residues are zero%
\begin{equation}
"N_{n}\left(  \xi_{j},\xi_{k}\right)  \ N_{n}\left(  \xi_{k},\xi_{j}\right)
"=H_{n}\left(  \xi_{j},\xi_{k}\right)  \ H_{n}\left(  \xi_{k},\xi_{j}\right)
+\frac{H_{n}\left(  \xi_{j},\xi_{k}\right)  -\ H_{n}\left(  \xi_{k},\xi
_{j}\right)  }{\xi_{j}-\xi_{k}}\tag{1.27}%
\end{equation}

\bigskip

In section 2, we develop the determinant $"D_{n}"$ in terms of fully packed
oriented loops configurations. Then, in section 3, we give an expression for
each configuration in the case of rotationnally invariant systems ; finally in
section 4, we use the Gaussian potential to obtain the correlation functions
(1.3) which in that case are nothing but the number of graphs obtained from
(1.6) by Wick's contraction. The large $N$ expansion (where $N$ is the size of
the matrix) is considered and provides the number of graphs according to their
genus (t'Hooft expansion $\left[  8\right]  $); another limit is also
considered, namely the so-called BMN limit $\left[  9\right]  $, where
$J_{i},$ $K_{i}\ $in (1.3) and $N\ $are large but the ratios $\frac{J_{i}%
}{\sqrt{N}}$ and $\frac{K_{i}}{\sqrt{N}}\ $are constant. Similar results were
presented in $\left[  10\right]  $ and $\left[  11\right]  .$

\bigskip

The end of this introduction is devoted to the description of the kernels
$K_{n},\ N_{n},\ A_{n}$ and to some of their properties. We refer to
Ref.$\left[  1\right]  $\ for the technical details; here, we simply give the
definition of these kernels$\ $%
\begin{equation}
K_{n}\left(  \xi,\eta^{\ast}\right)  =\sum_{i=0}^{n}\frac{p_{i}\left(
\xi\right)  \ p_{i}^{\ast}\left(  \eta\right)  }{h_{i}}\tag{1.28}%
\end{equation}
We introduce the following integrals (in the case of potentials of one
variable, the integral (1.29.a) is called the Cauchy-Hilbert transform of the
corresponding polynomial)%
\begin{align}
t_{n}\left(  y\right)   & =\int d^{2}z\ \ p_{n}^{\ast}\left(  z\right)
\ \ \frac{1}{z-y}\ \ \ e^{-V\left(  z,z^{\ast}\right)  }\tag{1.29.a}\\
Q\left(  x^{\ast},y\right)   & =\int d^{2}z\ \ \frac{1}{\left(  z^{\ast
}-x^{\ast}\right)  \left(  z-y\right)  }\ \ e^{-V\left(  z,z^{\ast}\right)
}\tag{1.29.b}%
\end{align}
then,%
\begin{align}
H_{n}\left(  \xi,y\right)   & =\sum_{i=0}^{n}\frac{p_{i}\left(  \xi\right)
\ t_{i}\left(  y\right)  }{h_{i}}\tag{1.30.a}\\
N_{n}\left(  \xi,y\right)   & =\frac{1}{y-\xi}+H_{n}\left(  \xi,y\right)
\tag{1.30.b}\\
A_{n}\left(  x^{\ast},y\right)   & =\sum_{i=0}^{n}\frac{t_{i}^{\ast}\left(
x\right)  \ t_{i}\left(  y\right)  }{h_{i}}-Q\left(  x^{\ast},y\right)
\tag{1.30.c}%
\end{align}
In (1.22), we also use the extensions
\begin{align}
K_{-1}\left(  \xi,\eta^{\ast}\right)   & =0\tag{1.31.a}\\
N_{-1}\left(  \xi,y\right)   & =\frac{1}{y-\xi}\tag{1.31.b}\\
A_{-1}\left(  x^{\ast},y\right)   & =-Q\left(  x^{\ast},y\right) \tag{1.31.c}%
\end{align}
and the determinant
\begin{align}
D_{-1}  & =\left\vert
\begin{array}
[c]{cc}%
N_{-1}\left(  \xi_{j},y_{i}\right)  & A_{-1}\left(  x_{l}^{\ast},y_{i}\right)
\\
0 & N_{-1}^{\ast}\left(  \eta_{k},x_{l}\right)
\end{array}
\right\vert \tag{1.32.a}\\
D_{-1}  & =\left\vert N_{-1}\left(  \xi_{j},y_{i}\right)  \right\vert
\ \ .\ \ \left\vert N_{-1}^{\ast}\left(  \eta_{k},x_{l}\right)  \right\vert
\tag{1.32.b}%
\end{align}
where%
\begin{align}
\left\vert N_{-1}\left(  \xi_{j},y_{i}\right)  \right\vert  & =\left\vert
\begin{array}
[c]{ccc}%
\frac{1}{y_{1}-\xi_{1}} & ... & \frac{1}{y_{1}-\xi_{M_{1}}}\\
... & ... & ...\\
\frac{1}{y_{M_{1}}-\xi_{1}} & ... & \frac{1}{y_{M_{1}}-\xi_{M_{1}}}%
\end{array}
\right\vert \tag{1.33.a}\\
\left\vert N_{-1}\left(  \xi_{j},y_{i}\right)  \right\vert  & =\left(
-\right)  ^{\frac{M_{1}\left(  M_{1}-1\right)  }{2}}\ \frac{\Delta\left(
y\right)  \Delta\left(  \xi\right)  }{%
{\displaystyle\prod\limits_{i,j}}
\left(  y_{i}-\xi_{j}\right)  }\tag{1.33.b}%
\end{align}

\bigskip

In Ref.$\left[  1\right]  $, we prove the following properties%
\begin{align}
\sum_{i=0}^{\infty}\frac{p_{i}\left(  \xi\right)  \ t_{i}\left(  y\right)
}{h_{i}}  & =\frac{1}{\xi-y}\tag{1.34.a}\\
\sum_{i=0}^{\infty}\frac{t_{i}^{\ast}\left(  x\right)  \ t_{i}\left(
y\right)  }{h_{i}}  & =Q\left(  x^{\ast},y\right) \tag{1.34.b}%
\end{align}
so that we have the formal power series%
\begin{align}
N_{n}\left(  \xi,y\right)   & =-\sum_{i=n+1}^{\infty}\frac{p_{i}\left(
\xi\right)  \ t_{i}\left(  y\right)  }{h_{i}}\tag{1.35.a}\\
A_{n}\left(  x^{\ast},y\right)   & =-\sum_{i=n+1}^{\infty}\frac{t_{i}^{\ast
}\left(  x\right)  \ t_{i}\left(  y\right)  }{h_{i}}\tag{1.35.b}%
\end{align}
Finally, we have the asymptotic behaviour%
\begin{align}
t_{n}\left(  y\right)   & \sim-\frac{h_{n}}{y^{n+1}}\ \ \text{as}%
\ y\rightarrow\infty\tag{1.36.a}\\
N_{n}\left(  \xi,y\right)   & \sim\frac{p_{n+1}\left(  \xi\right)  }{y^{n+2}%
}\ \ \text{as}\ y\rightarrow\infty\tag{1.36.b}\\
A_{n}\left(  x^{\ast},y\right)   & \sim\frac{t_{n+1}^{\ast}\left(  x\right)
}{y^{n+2}}\ \ \text{as}\ y\rightarrow\infty\tag{1.36.c}%
\end{align}

\bigskip

\section{\bigskip\bigskip The graph description of the determinant "$D_{n}$"}

The determinant "$D_{n}$"\ as given in (1.26) is a sum of products of oriented
propagators joining the various points (for simplicity, we label the points
and the corresponding variables by the same letter)$.$ Let us split the plane
into two parts separated by a border line; on the right part we draw the
various points $\eta_{i}^{\ast}$ and in the left part we draw the various
points $\xi_{i}$. As we develop the determinant, every time we meet a kernel
$A_{n}\left(  \eta_{i}^{\ast},\xi_{j}\right)  \ $we draw an oriented
propagator from the point $\eta_{i}^{\ast}$ towards the point $\xi_{j}%
\ $crossing the border from right to left; similarly with the kernel
$K_{n}\left(  \xi_{i},\eta_{j}^{\ast}\right)  $ we draw an oriented propagator
from the point $\xi_{i}$ towards the point $\eta_{j}^{\ast}\ $crossing the
border from left to right.\ Then, inside the $\xi$-region we draw an oriented
propagator from the point $\xi_{i}$ towards the point $\xi_{j}\ $corresponding
to the kernel $N_{n}\left(  \xi_{i},\xi_{j}\right)  $; in addition, the
diagonal part of $"D_{n}"$ defines self closing loops on one point $\xi_{i}$
from the kernel $H_{n}\left(  \xi_{i},\xi_{i}\right)  $.\ Similarly, inside
the $\eta^{\ast}$-region we draw an oriented propagator from the point
$\eta_{i}^{\ast}$ towards the point $\eta_{j}^{\ast}\ $corresponding to the
kernel $N_{n}^{\ast}\left(  \eta_{j},\eta_{i}\right)  $; in addition, the
diagonal part of $"D_{n}"$ defines self closing loops on one point $\eta
_{i}^{\ast}$ from the kernel $H_{n}^{\ast}\left(  \eta_{i},\eta_{i}\right)  $.\ 

As we develop the determinant, the product of propagators describes a number
of disjoint oriented closed loops $\gamma_{a}\ $(including self-loop with one
point) so that any point $\xi_{i}$ or $\eta_{i}^{\ast}$ belongs to one closed
loop $\gamma_{a}$ and only one; we call such a configuration "fully packed
oriented loops configuration" which we denote by $\varpi$\ (fig.1); the number
of oriented loops in$\ \varpi$ is denoted $\#\left(  \varpi\right)  $%
\begin{equation}
\varpi=\cup_{a=1}^{\#\left(  \varpi\right)  }\gamma_{a}\tag{2.1}%
\end{equation}
%

\begin{center}
\includegraphics[
height=6.882cm,
width=10.1067cm
]%
{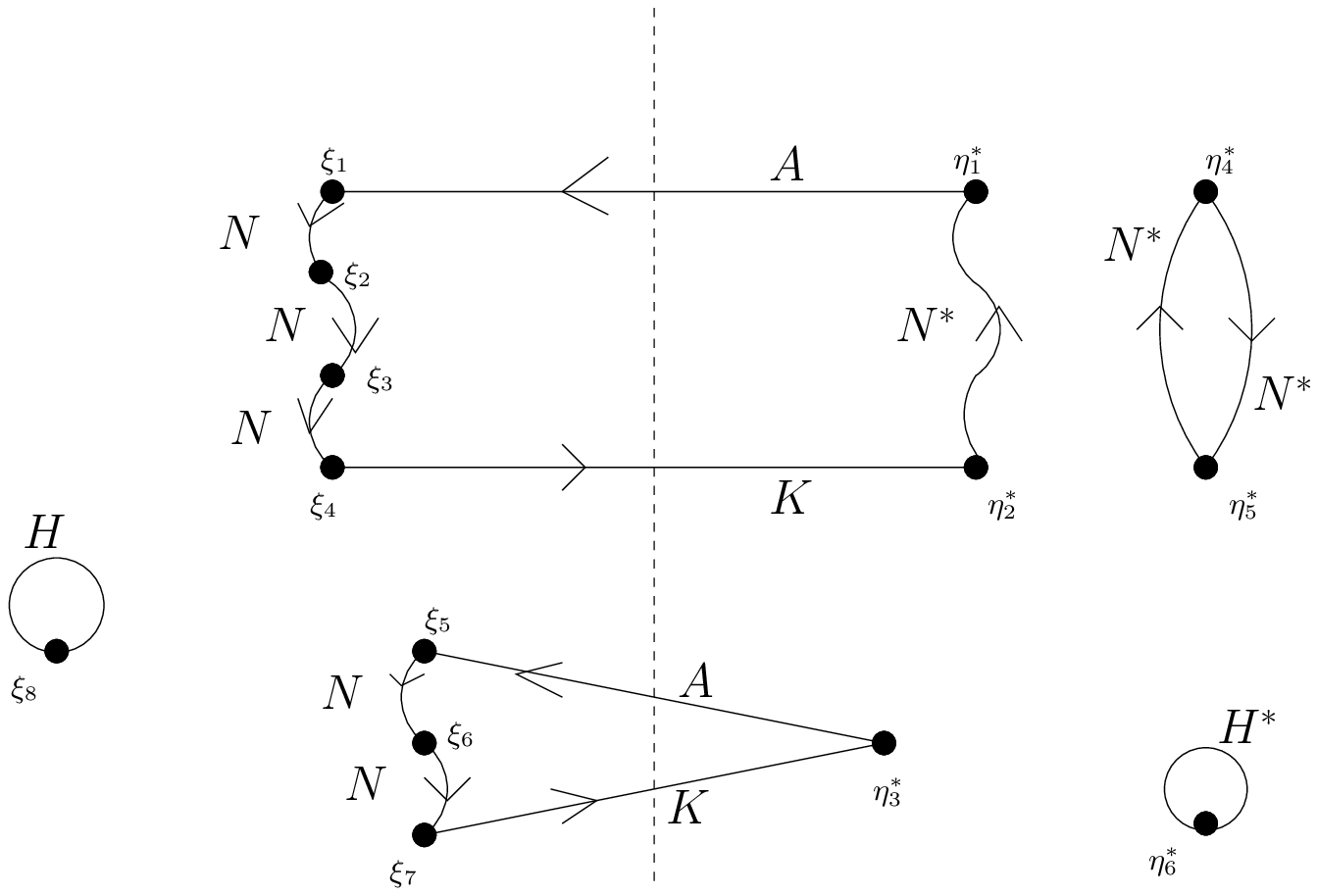}%
\\
fig.1 : contribution to $W_{8,6}\left(  \xi_{1},\ldots,\xi_{8};\eta_{1}%
^{\star},\ldots,\eta_{6}^{\star}\right)  $%
\label{fig001}%
\end{center}

\bigskip

\bigskip

The determinant "$D_{n}$" is the sum over the $\left(  M_{1}+M_{2}\right)
!\ $configurations $\varpi$%
\begin{equation}
"D_{n}"=\left(  -\right)  ^{M_{1}+M_{2}}\ \sum_{\varpi}\
{\displaystyle\prod\limits_{a=1}^{\#\left(  \varpi\right)  }}
\left(  -I_{\gamma_{a}}\right)  \ \ \tag{2.2}%
\end{equation}

\bigskip

Given an oriented loop $\gamma,$ which is not made of $N$ propagators alone
and/or of $N^{\ast}$ propagators alone, the amplitude $I_{\gamma}$ is the
product of a certain number $q>0$ of propagators $K_{n}$ and of the same
number $q$ of propagators $A_{n};$ the remaining propagators describe $q$
chains of propagators $N$ and $q$ chains of propagators $N^{\ast}\ $(such
chains may be reduced to a single point).$\ $We call skeleton graph an
oriented loop $\gamma$ without $N$ and $N^{\ast}$ propagators. To any oriented
loop $\gamma\ $we associate a skeleton graph by contraction of each of its
chains of propagators $N$ or $N^{\ast}$ into a point.

Along a given chain, the propagators $N$ or $N^{\ast}$ have poles when the
$\xi^{\prime}s$ or the $\eta^{\prime}s$ coincide. However, these poles cancel
when we sum over all permutations of the $\xi^{\prime}s$ or of the
$\eta^{\prime}s$ inside a given chain of propagators $N$ or $N^{\ast}.$ In
appendix A we show how the poles cancel when we sum over several
configurations which have the same associated skeleton graph.

\bigskip

For a given potential $V\left(  z,z^{\ast}\right)  ,\ $we must calculate the
orthogonal polynomials $p_{n}\left(  z\right)  $ and their Cauchy transform
$t_{n}\left(  z\right)  $ in order to evaluate the kernels which are the
propagators of the graphs. Of course, each case is a specific case; however,
some general structure can be obtained in the special case where the system
(potential and domain of integration) is rotationnally invariant. $\ $

\bigskip

\bigskip

\section{The $\frac{1}{\xi},\ \frac{1}{\eta^{\ast}}\ $expansion in the case of
the rotationnally invariant systems}

\bigskip

The expressions for the four kernels $K_{n},\ N_{n},\ N_{n}^{\ast},A_{n}$
simplify greatly in the case of rotationnally invariant systems because the
angular integrations in (1.17) is trivially performed. Given a potential
$V\left(  zz^{\ast}\right)  \ $and a disc of radius $R$ which might be the
complex plane ($R=\infty)\ $if $\exp\left(  -V\left(  zz^{\ast}\right)
\right)  $ decreases strongly enough at $\infty$.\ In that case, the
orthogonal polynomials are the so-called Ginibre's polynomials%
\begin{equation}
p_{n}\left(  z\right)  =z^{n}\tag{3.1}%
\end{equation}
\bigskip and their norms squared are%
\begin{equation}
h_{n}=2\pi\int_{0}^{R}d\rho\ \rho^{2n+1}\ e^{-V\left(  \rho^{2}\right)
}\tag{3.2}%
\end{equation}
In the Gaussian case where $V\left(  zz^{\ast}\right)  =zz^{\ast}$ and
$R=\infty,$%
\begin{equation}
h_{n}=\pi\ \Gamma\left(  n+1\right) \tag{3.3}%
\end{equation}
The function $t_{n}\left(  y\right)  $ defined in (1.29.a) is found to be%
\begin{equation}
t_{n}\left(  y\right)  =-\frac{h_{n}}{y^{n+1}},\ \ \ \ \ \ \ \ \left\vert
y\right\vert >R\ \tag{3.4}%
\end{equation}
if $R=\infty$ this result is still valid up to exponentially small terms when
$y\rightarrow\infty$ and can be used as a formal power series (see also (1.34.a)).

\bigskip

\bigskip According to (1.28) and (1.30)\ the formal power series for the kernels

$K_{n},\ H_{n},N_{n}$ and $A_{n}$ are%
\begin{align}
K_{n}\left(  \xi,\eta^{\ast}\right)   & =\sum_{j=0}^{n}\frac{\left(  \xi
\eta^{\ast}\right)  ^{j}}{h_{j}}\tag{3.5.a}\\
H_{n}\left(  \xi_{i},\xi_{j}\right)   & =-\frac{1}{\xi_{j}^{n+1}}\ \frac
{\xi_{i}^{n+1}-\xi_{j}^{n+1}}{\xi_{i}-\xi_{j}}\ \ \ \ \ \xi_{i}\neq\xi
_{j}\tag{3.5.b}%
\end{align}
so that%
\begin{align}
H_{n}\left(  \xi_{i},\xi_{i}\right)   & =-\frac{n+1}{\xi_{i}}\tag{3.6.a}\\
N_{n}\left(  \xi_{i},\xi_{j}\right)   & =\left(  \frac{\xi_{i}}{\xi_{j}%
}\right)  ^{n+1}\frac{1}{\xi_{j}-\xi_{i}}\tag{3.6.b}%
\end{align}
Finally,%

\begin{equation}
A_{n}\left(  \eta^{\ast},\xi\right)  =-\sum_{i=n+1}^{\infty}\frac{h_{i}%
}{\left(  \xi\eta^{\ast}\right)  ^{i+1}}\tag{3.7}%
\end{equation}

It is now possible to calculate for any closed loop $\gamma$ the corresponding
amplitude $I_{\gamma}\ $which contributes to the value of the determinant
\bigskip$"D_{n}".$

\bigskip

\textbf{1}$%
{{}^\circ}%
)\ $\textbf{the }$\frac{1}{\xi},\ \frac{1}{\eta^{\ast}}\ $\textbf{expansion
for the skeleton graphs:\ }given a skeleton graph with $q>0$ propagators $A$
and $q$ propagators $K$, the expression for $I_{\gamma}\ $can be written%
\begin{equation}
I_{\gamma}=%
{\displaystyle\prod\limits_{k=1}^{q}}
A_{n}\left(  \eta_{\alpha_{k}}^{\ast},\xi_{\alpha_{k}}\right)  \ \
{\displaystyle\prod\limits_{i}^{q}}
K_{n}\left(  \xi_{\alpha_{k}},\eta_{\alpha_{k+1}}^{\ast}\right)
\ \ \ \ ,\text{with}\ \ \ \eta_{\alpha_{q+1}}^{\ast}=\eta_{\alpha_{1}}^{\ast
}\tag{3.8}%
\end{equation}
As we expand the propagators $A_{n}$ and $K_{n}$ according to (3.5.a) and
(3.7) we introduce an internal momentum $i_{k}$ for $A_{n}\left(  \eta
_{\alpha_{k}}^{\ast},\xi_{\alpha_{k}}\right)  $ and an internal momentum
$j_{k}$ for $K_{n}\left(  \xi_{\alpha_{k}},\eta_{\alpha_{k+1}}^{\ast}\right)
;\ $the summation over $i_{k}\ $runs from $n+1\ $to $\infty$ while the
summation over $j_{k}$ runs from $0$ to $n.\ $By convention, we extend the
definition of $h_{j}$ to negative $j^{\prime}s$ as $h_{j<0}=\infty$ so that in
$K_{n}$ we may eventually ignore the lower limit of summation.

The expansion of $I_{\gamma}$ is of the form%
\begin{equation}
I_{\gamma}=\sum_{\left\{  J^{\prime}s>0,K^{\prime}s>0\right\}  }\
{\displaystyle\prod\limits_{k=1}^{q}}
\left(  \frac{1}{\xi_{\alpha_{k}}^{J_{k}+1}}\right)  \
{\displaystyle\prod\limits_{k=1}^{q}}
\left(  \frac{1}{\eta_{\alpha_{k}}^{K_{k}+1}}\right)  ^{\ast}\ \ \ I\left(
J_{k},K_{k}\right) \tag{3.9}%
\end{equation}
where
\begin{equation}
I\left(  J_{k},K_{k}\right)  =\left(  -\right)  ^{q}\sum_{D_{\left\{
J_{k},K_{k}\right\}  }}\ \frac{h_{i_{1}}...h_{i_{q}}}{h_{j_{1}}...h_{jq}%
}\tag{3.10}%
\end{equation}
and $D_{\left\{  J_{k},K_{k}\right\}  }$ is the summation domain for the
indices $i_{k}$ and $j_{k}.$ The determination of this domain is the main
difficulty. Clearly, all the indices $i^{\prime}s$ and $j^{\prime}s$ are fixed
in terms of the $J^{\prime}s$ and the $K^{\prime}s$ up to a translation.\ 

\bigskip\ We first describe some simple exemples:

a$)\ $ let us consider the graph $\gamma_{1}$ (fig.2).%

\begin{center}
\includegraphics[
height=5.5992cm,
width=7.0973cm
]%
{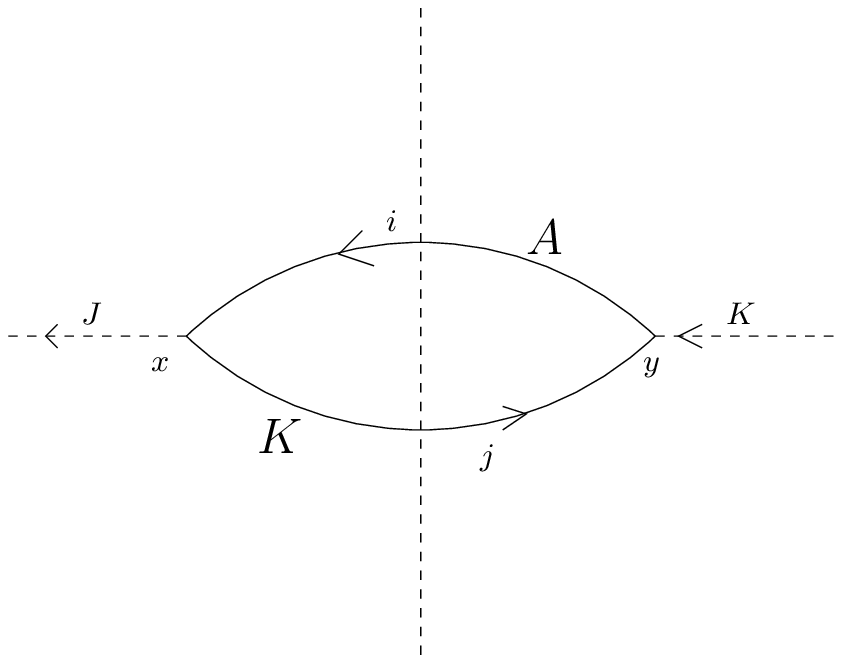}%
\\
fig. 2 : graph $\gamma_{1}\rightarrow W_{1,1}^{c}\left(  x,y\right)  $%
\label{fig002}%
\end{center}

In that case, the momentum conservation law is
\begin{equation}
i-j=J=K\tag{3.11}%
\end{equation}
so that the domain of summation for the momentum $j$ is%
\begin{equation}
n-J+1\leq j\leq n\tag{3.12}%
\end{equation}
Consequently,%
\begin{equation}
I_{\gamma_{1}}\left(  J,K\right)  =-\delta_{J,K}\ \ \sum_{j=n-J+1}^{n}%
\frac{h_{j+J}}{h_{j}}\tag{3.13}%
\end{equation}
The equation (3.13) is still valid in the case where $n-J+1<0$ with the
convention $h_{j<0}=\infty.$\ We define the characteristic function for the
skeleton graph $\gamma_{1}$%
\begin{equation}
\Phi_{\gamma_{1}}\left(  n,J\right)  =\ \sum_{j=0}^{n}\frac{h_{j+J}}{h_{j}%
}\tag{3.14}%
\end{equation}
then,%
\begin{equation}
I_{\gamma_{1}}\left(  J,K\right)  =-\delta_{J,K}\ \ \left(  1-T_{J}\right)
\ \Phi_{\gamma_{1}}\left(  n,J\right) \tag{3.15}%
\end{equation}
where the operator $T_{J}$ is a translation operator over $n$%
\begin{equation}
T_{J}\ f\left(  n\right)  =f\left(  n-J\right) \tag{3.16}%
\end{equation}
\pagebreak

b$)\ $let us consider the graph $\gamma_{2}$ (fig.3).%

\begin{center}
\includegraphics[
height=4.257cm,
width=7.0973cm
]%
{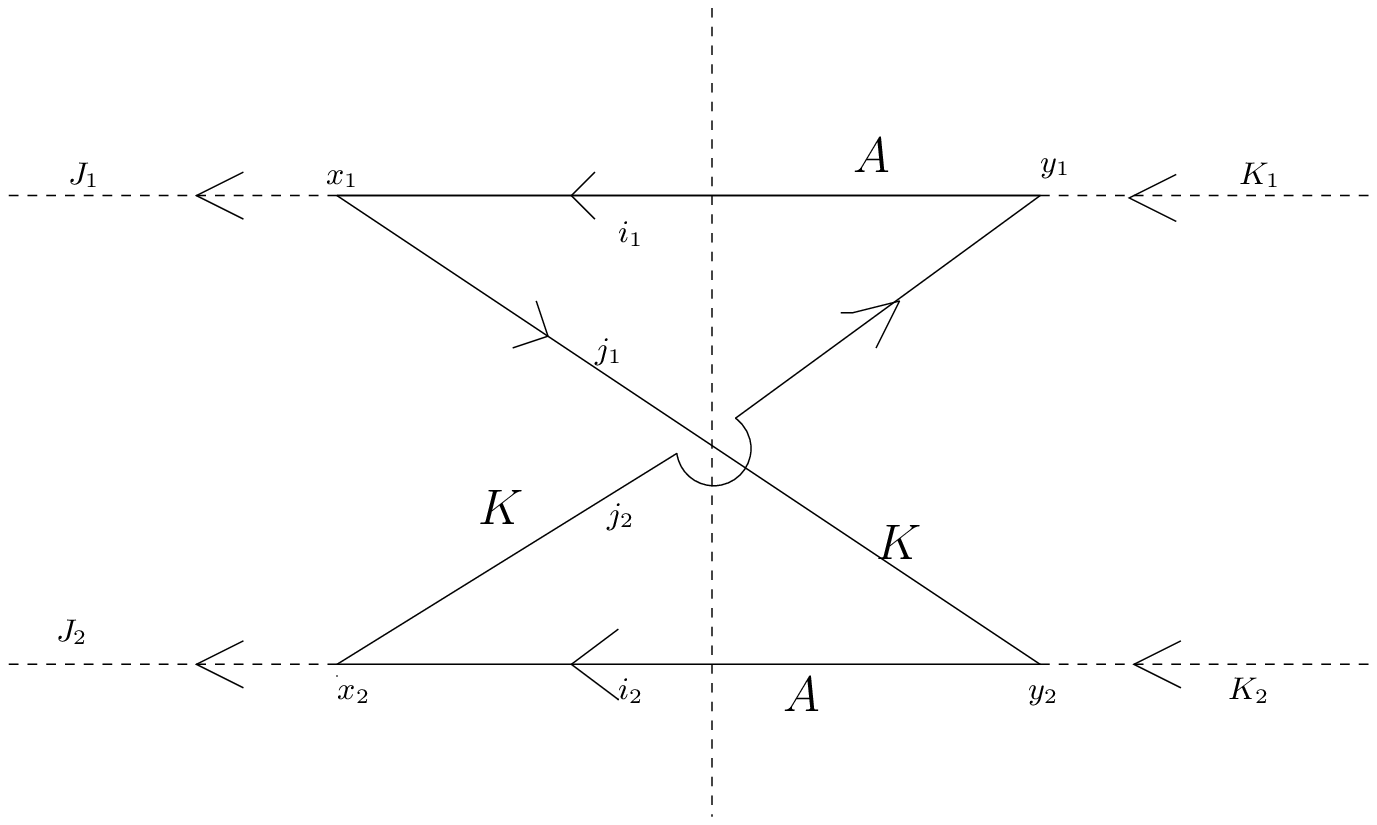}%
\\
fig. 3 : graph $\gamma_{2}\rightarrow$ contribution to $W_{2,2}^{c}\left(
x_{1},x_{2};y_{1},y_{2}\right)  $%
\label{fig003}%
\end{center}

Choosing\ $j_{2}=j$ as the independant loop momentum, we observe the following
constraints%
\begin{align}
j  & \geq n+1-K_{1},\ \ \ \ \ \ j\geq n+1-J_{2}\ \ \tag{3.17.a}\\
j  & \leq n-K_{1}+J_{1},\ \ \ \ j\leq n\tag{3.17.b}\\
J  & =J_{1}+J_{2}=K_{1}+K_{2}=K\tag{3.17.c}%
\end{align}
Consequently, we obtain%
\begin{align}
I_{\gamma_{2}}\left(  J_{1},J_{2},K_{1},K_{2}\right)   & =\delta
_{J,K}\ O\left(  J_{i},K_{i}\right)  \ \Phi_{\gamma_{2}}\left(  n,J_{i}%
,K_{i}\right) \tag{3.18.a}\\
O\left(  J_{i},K_{i}\right)   & =T_{Sup\left(  K_{1}-J_{1},0\right)
}-T_{Inf\left(  K_{1},J_{2}\right)  }\tag{3.18.b}\\
\Phi_{\gamma_{2}}\left(  n,J_{i},K_{i}\right)   & =\ \sum_{j=0}^{n}%
\frac{h_{j+K_{1}}\ \ \ \ h_{j+J_{2}}}{h_{j+K_{1}-J_{1}}\ \ \ h_{j}%
}\tag{3.18.c}%
\end{align}

\bigskip

\bigskip

c$)$\ the generalization to the skeleton graph $\gamma_{q}$ with $q$ pairs of
propagators $A_{n}$ and $K_{n}\ $is now straightforward. We define%
\begin{equation}
J=\sum_{k=1}^{q}J_{k}\ \ \ \ \ \ \ K=\sum_{k=1}^{q}K_{k}\tag{3.19}%
\end{equation}
and we choose $j=j_{q}$ as independant loop momentum so that%
\begin{equation}
i_{k}=j_{q}+b_{k}\ \ \ \ \ \ \ \ \ \ \ \ \ \ \ j_{k}=j_{q}+a_{k}\tag{3.20}%
\end{equation}
with%
\begin{align}
a_{k}  & =\sum_{p=1}^{k}\left(  K_{p}-J_{p}\right)  ,\ \ \ \ \ \ \ a_{q=0}%
\tag{3.21.a}\\
b_{k}  & =a_{k}+J_{k}\geq1\tag{3.21.b}%
\end{align}
We note from (3.20) that any variable $b_{k}$ is strictly larger than any
variable $a_{k^{\prime}}.$ The contribution of the graph $\gamma_{q}$ is
\begin{align}
I_{\gamma_{q}}\left(  J_{i},K_{i}\right)   & =\left(  -\right)  ^{q}%
\ \delta_{J,K}\ \left(  T_{X}-T_{Y}\right)  \ \Phi_{\gamma_{q}}\left(
n,J_{i},K_{i}\right) \tag{3.22.a}\\
\Phi_{\gamma_{q}}\left(  n,J_{i},K_{i}\right)   & =\sum_{j=0}^{n}%
\ \frac{h_{j+b_{1}}\ ....h_{j+b_{q}}}{h_{j+a_{1}}\ ....h_{j+a_{q}}%
}\tag{3.22.b}%
\end{align}
with%
\begin{align}
0  & \leq X=Sup\left(  a_{k}\right) \tag{3.23.a}\\
1  & \leq Y=Inf\left(  b_{k}\right) \tag{3.23.b}%
\end{align}

\bigskip

\bigskip

\textbf{2}$%
{{}^\circ}%
)$\textbf{\ the }$\frac{1}{\xi},\ \frac{1}{\eta}\ $\textbf{expansion for any
graph:\ }a one loop graph $\gamma$ is made of $q$ propagators $A,$ $q$
propagators $K$, $q$ chains $\left(  C_{1},...,C_{q}\right)  \ $of propagators
$N$ and $q$ chains $\left(  C_{1}^{\ast},...,C_{q}^{\ast}\right)  \ $of
propagators\ $N^{\ast}\ $(the chains can be reduced to a single point). The
case $q=0$ is made\ of loops of $N$ propagators alone, or of $N^{\ast}$
propagators alone. To each chain $C_{i}\ $we associate an external outgoing
momentum%
\begin{equation}
\Delta_{i}=\sum_{\xi_{k}\in C_{i}}\ J_{k}\tag{3.24}%
\end{equation}
and to each chain $C_{i}^{\ast}\ $an external ingoing momentum%
\begin{equation}
\Theta_{i}=\sum_{\eta_{k}\in C_{i}^{\ast}}\ K_{k}\tag{3.25}%
\end{equation}
Now, to each graph $\gamma$ we associate a skeleton graph $s\left(
\gamma\right)  $ obtained from $\gamma$ by shrinking into a point each of the
chain $C_{i}\ $or $C_{i}^{\ast}$. To each of these points we associate the
corresponding external momentum $\Delta_{i}$ or $\Theta_{i};$ we have
\begin{equation}
J=\sum_{i=1}^{q}\Delta_{i}=\sum_{i=1}^{q}\Theta_{i}=\sum_{\xi_{k}\in\gamma
}J_{k}=\sum_{\eta_{k}\in\gamma}K_{k}=K\tag{3.26}%
\end{equation}
Because all $J^{\prime}s$ and $K^{\prime}s$ are positive or nul, equation
(3.26) shows that if the graph $\gamma\ $is made of $N$ propagators alone
$(q=0),\ $since the $K^{\prime}s$ are zero, the $J^{\prime}s$ are also zero.
The only non zero contribution in this category is the self-closed loop with
one point which is $H_{n}\left(  \xi_{i},\xi_{i}\right)  =-\frac{n+1}{\xi_{i}%
}.\ $The two points loop is%
\begin{equation}
"N\left(  \xi_{i},\xi_{j}\right)  \ N\left(  \xi_{j},\xi_{i}\right)
"=0\tag{3.27}%
\end{equation}
Larger loops of $N$ propagators alone, or of $N^{\ast}$ propagators alone are
shown in appendix A to be zero.

The functions $h_{i\text{ }}$are only present in the $A$ and $K$ propagators
so that the function to be summed for a graph $\gamma\ $is the characteristic
function of the skeleton graph $s\left(  \gamma\right)  $%
\begin{equation}
\frac{h_{j+b_{1}}...h_{j+b_{q}}}{h_{j+a_{1}}...h_{j+a_{q}}}\tag{3.28}%
\end{equation}
$\ $where $j$ is the momentum corresponding to the propagator $K_{q}\ $and%
\begin{align}
a_{k}  & =\sum_{p=1}^{k}\left(  \Theta_{p}-\Delta_{p}\right)
,\ \ \ \ \ \ \ a_{q=0}\tag{3.29.a}\\
b_{k}  & =a_{k}+\Delta_{k}\geq1\tag{3.29.b}%
\end{align}
The equation (3.10) for a one loop graph $\gamma$ with propagators $N$ and
$N^{\ast}$ can be written
\begin{equation}
I_{\gamma}\left(  J_{i},K_{i}\right)  =\left(  -\right)  ^{q}\ \delta
_{J,K}\ O\left(  J_{i},K_{i}\right)  \ \sum_{j=0}^{n}\ \frac{h_{j+b_{1}%
}...h_{j+b_{q}}}{h_{j+a_{1}}...h_{j+a_{q}}}\tag{3.30}%
\end{equation}
where the operator $O\left(  J_{i},K_{i}\right)  $ determins the range of
summation for the variable $j$ and is described below.

We introduce the commutative operation $\otimes$%
\begin{align}
T_{a}\otimes T_{b}  & =T_{Sup\left(  a,b\right)  }\tag{3.31.a}\\
1\otimes T_{b}  & =T_{Sup\left(  0,b\right)  }\tag{3.31.b}%
\end{align}
so that conditions like $n-a+1\leq j\leq n$ together with $n-b+1\leq j\leq n,
$ where $a$ and $b$ are positive, can be written%
\begin{equation}
\sum_{j=n-Inf\left(  a,b\right)  +1}^{n}...=\left(  1-T_{a}\right)
\otimes\left(  1-T_{b}\right)  \ \sum_{j=0}^{n}...\tag{3.32}%
\end{equation}
Now, we proved in appendix A that for a given chain $C_{i}\ $with $p$ points
($\left(  p-1\right)  $ propagators), the sum over the $p!$ permutations
generates a domain of summation for the variable $j_{i}\ $described by the
operator $%
{\displaystyle\prod\limits_{\xi_{r}\in C_{i}}}
\left(  1-T_{J_{r}}\right)  ;$ since the propagator with momentum $j_{i}$ is
between the chains $C_{i}$ and $C_{i+1}^{\ast}$ (with $C_{q+1}^{\ast}%
=C_{1}^{\ast}$) the constraints on $j_{i}$ can be written%
\begin{equation}
\left[
{\displaystyle\prod\limits_{\xi_{r}\in C_{i}}}
\left(  1-T_{J_{r}}\right)  \right]  \otimes\left[
{\displaystyle\prod\limits_{\eta_{s}\in C_{i+1}^{\ast}}}
\left(  1-T_{K_{s}}\right)  \right] \tag{3.33}%
\end{equation}
and using the relation (3.20) we can write the corresponding constraints on
$j_{q}\ $as%
\begin{equation}
T_{a_{i}}\left\{  \left[
{\displaystyle\prod\limits_{\xi_{r}\in C_{i}}}
\left(  1-T_{J_{r}}\right)  \right]  \otimes\left[
{\displaystyle\prod\limits_{\eta_{s}\in C_{i+1}^{\ast}}}
\left(  1-T_{K_{s}}\right)  \right]  \right\} \tag{3.34}%
\end{equation}
We collect all the constraints on the various momenta $j_{i}\ $for $i=1,...,q
$ and write%
\begin{equation}
O\left(  J_{i},K_{i}\right)  =\otimes_{i=1}^{q}\left[  T_{a_{i}}\left\{
\left[
{\displaystyle\prod\limits_{\xi_{r}\in C_{i}}}
\left(  1-T_{J_{r}}\right)  \right]  \otimes\left[
{\displaystyle\prod\limits_{\eta_{s}\in C_{i+1}^{\ast}}}
\left(  1-T_{K_{s}}\right)  \right]  \right\}  \right] \tag{3.35}%
\end{equation}

\bigskip

\section{\bigskip The Gaussian potential, large $N$ and BMN expansion}

We now apply the results of section 3 to the Gaussian potential with the
function $h_{n}$ given in (3.3); we also replace the index $n$ of section 3 by
$\left(  N-1\right)  $ according to (1.25).

1$%
{{}^\circ}%
)\ $let us consider the two points function%
\begin{equation}
W_{1,1}\left(  N,x,y\right)  =\ <Tr\frac{1}{x-M}\ Tr\frac{1}{y-M^{+}%
}>\tag{4.1}%
\end{equation}
which is equal from (1.26) to the determinant%
\begin{equation}
"D_{N-1}"="\det"\left\vert
\begin{array}
[c]{cc}%
H_{N-1}\left(  x,x\right)  & A_{N-1}\left(  y,x\right) \\
K_{N-1}\left(  x,y\right)  & H_{N-1}\left(  y,y\right)
\end{array}
\right\vert \tag{4.2}%
\end{equation}
The part $H_{N-1}\left(  x,x\right)  \ H_{N-1}\left(  y,y\right)  $ of the
determinant corresponds to a disconnected graph made of two self closing loops
and from (3.6.a) is equal to$\ \frac{N^{2}}{xy}$. The part $A_{N-1}\left(
y,x\right)  \ K_{N-1}\left(  x,y\right)  $ of the determinant corresponds to
the graph $\gamma_{1}\ $(fig.2). The corresponding function $\Phi_{\gamma_{1}%
}\left(  N-1,J\right)  \ $is given in (3.14)%
\begin{equation}
\Phi_{\gamma_{1}}\left(  N-1,J\right)  =\sum_{j=0}^{N-1}\frac{\Gamma\left(
j+J+1\right)  }{\Gamma\left(  j+1\right)  }=\frac{1}{J+1}\ \frac{\Gamma\left(
N+J+1\right)  }{\Gamma\left(  N\right)  }\tag{4.3}%
\end{equation}
Consequently, according to (3.15)%
\begin{equation}
A_{N-1}\left(  y,x\right)  \ K_{N-1}\left(  x,y\right)  =-\sum_{J=1}^{\infty
}\frac{1}{\left(  xy\right)  ^{J+1}}\ \frac{1}{J+1}\left[  \frac{\Gamma\left(
N+J+1\right)  }{\Gamma\left(  N\right)  }-\frac{\Gamma\left(  N+1\right)
}{\Gamma\left(  N-J\right)  }\right] \tag{4.4}%
\end{equation}
The square bracket $\left[  \ \ \right]  $ in (4.4) has clearly a $\frac
{1}{N^{2}}$ expansion
\begin{equation}
\frac{\Gamma\left(  N+J+1\right)  }{\Gamma\left(  N\right)  }-\frac
{\Gamma\left(  N+1\right)  }{\Gamma\left(  N-J\right)  }=2\ N^{J}%
\ \sum_{p\ odd>0}\ \frac{1}{N^{p-1}}\ \sigma_{p}\left(  J\right) \tag{4.5}%
\end{equation}
where $\sigma_{p}\left(  J\right)  $ is described in appendix B. After a
convenient rescaling, we write $A_{N-1}\left(  y\sqrt{N},x\sqrt{N}\right)
\ K_{N-1}\left(  x\sqrt{N},y\sqrt{N}\right)  $ as$\ $
\begin{align}
& -\frac{1}{N}\sum_{J=1}^{\infty}\frac{1}{\left(  xy\right)  ^{J+1}}\left[
J+\frac{1}{24N^{2}}\left(  J+1\right)  J^{2}\left(  J-1\right)  \left(
J-2\right)  +O\left(  \frac{1}{N^{4}}\right)  \right] \nonumber\\
& \tag{4.6.a}\\
& =-\frac{1}{N}\left[  \frac{1}{\left(  xy-1\right)  ^{2}}+\frac{1}{N^{2}%
}\ \frac{xy\left(  3xy+2\right)  }{\left(  xy-1\right)  ^{6}}+O\left(
\frac{1}{N^{4}}\right)  \right] \tag{4.6.b}%
\end{align}

\bigskip

The so-called BMN limit of $I_{\gamma_{1}}\left(  J,K\right)  \ $(defined in
(3.15), (4.3)) is the large ~$N,\ $large $J$ limit with constant $\frac{J^{2}%
}{N},\ \frac{K^{2}}{N}\ .\ $From the large $J$ behaviour of$\ \sigma
_{p}\left(  J\right)  $ as given in appendix B, we obtain the BMN limit as%
\begin{equation}
I_{\gamma_{1}}\left(  J,K\right)  \sim-\delta_{J,K}\ \ \frac{2\ N^{J+1}}%
{J}\ sh\left(  \frac{J^{2}}{2N}\right) \tag{4.7}%
\end{equation}

\bigskip

At this stage, we find convenient to introduce a formalism developped in
appendix B and which makes the $\frac{1}{N^{2}\ }$expansion and the BMN limit
transparant. We define%
\begin{equation}
N^{\ast\ast(J+1)}=\left(  N-\frac{J}{2}\right)  \left(  N-\frac{J}%
{2}+1\right)  ...\left(  N+\frac{J}{2}-1\right)  \left(  N+\frac{J}{2}\right)
\tag{4.8}%
\end{equation}
which has manifestly a $\frac{1}{N^{2}\ }$expansion%
\begin{equation}
N^{\ast\ast(J+1)}=\sum_{p=0}^{E\left(  \frac{J+1}{2}\right)  }\left(
-\right)  ^{p}\ \Sigma_{2p}\left(  J\right)  \ N^{J-2p+1}\tag{4.9}%
\end{equation}
where the coefficients $\ \Sigma_{2p}\left(  J\right)  $ are given in appendix
B. Consequently, the expression (4.4) can be written as%
\begin{equation}
\left(  1-T_{J}\right)  \ \frac{1}{J+1}\ \frac{\Gamma\left(  N+J+1\right)
}{\Gamma\left(  N\right)  }=\frac{2}{J+1}\ sh\left(  \frac{J}{2}%
\ \frac{\partial}{\partial N}\right)  \ N^{\ast\ast(J+1)}\tag{4.10}%
\end{equation}
which is manifestly a $\frac{1}{N^{2}\ }$expansion.\ Also the BMN limit of
$N^{\ast\ast(J+1)}$ is simply $N^{J+1}$ so that the BMN limit of (4.10)
consists in replacing $\frac{\partial}{\partial N}$ by $\frac{J}{N}\ $and
$N^{\ast\ast(J+1)}$ by $N^{J+1}.$

\bigskip

2$%
{{}^\circ}%
)\ $we now consider the three points function%
\begin{equation}
W_{2,1}\left(  N,x_{1},x_{2},y\right)  =\ <Tr\frac{1}{x_{1}-M}\ Tr\frac
{1}{x_{2}-M}Tr\frac{1}{y-M^{+}}>\tag{4.11}%
\end{equation}
which is equal from (1.26) to the determinant%
\begin{equation}
"D_{N-1}"="\det"\left\vert
\begin{array}
[c]{ccc}%
H_{N-1}\left(  x_{1},x_{1}\right)  & N_{N-1}\left(  x_{2},x_{1}\right)  &
A_{N-1}\left(  y,x_{1}\right) \\
N_{N-1}\left(  x_{1},x_{2}\right)  & H_{N-1}\left(  x_{2},x_{2}\right)  &
A_{N-1}\left(  y,x_{2}\right) \\
K_{N-1}\left(  x_{1},y\right)  & K_{N-1}\left(  x_{2},y\right)  &
H_{N-1}\left(  y,y\right)
\end{array}
\right\vert \tag{4.12}%
\end{equation}
The connected contribution to (4.12) is

$\left[  A_{N-1}\left(  y,x_{1}\right)  \ N_{N-1}\left(  x_{1},x_{2}\right)
\ K_{N-1}\left(  x_{2},y\right)  +\left(  x_{1}\Leftrightarrow x_{2}\right)
\right]  .$%

\begin{center}
\includegraphics[
height=4.257cm,
width=7.0951cm
]%
{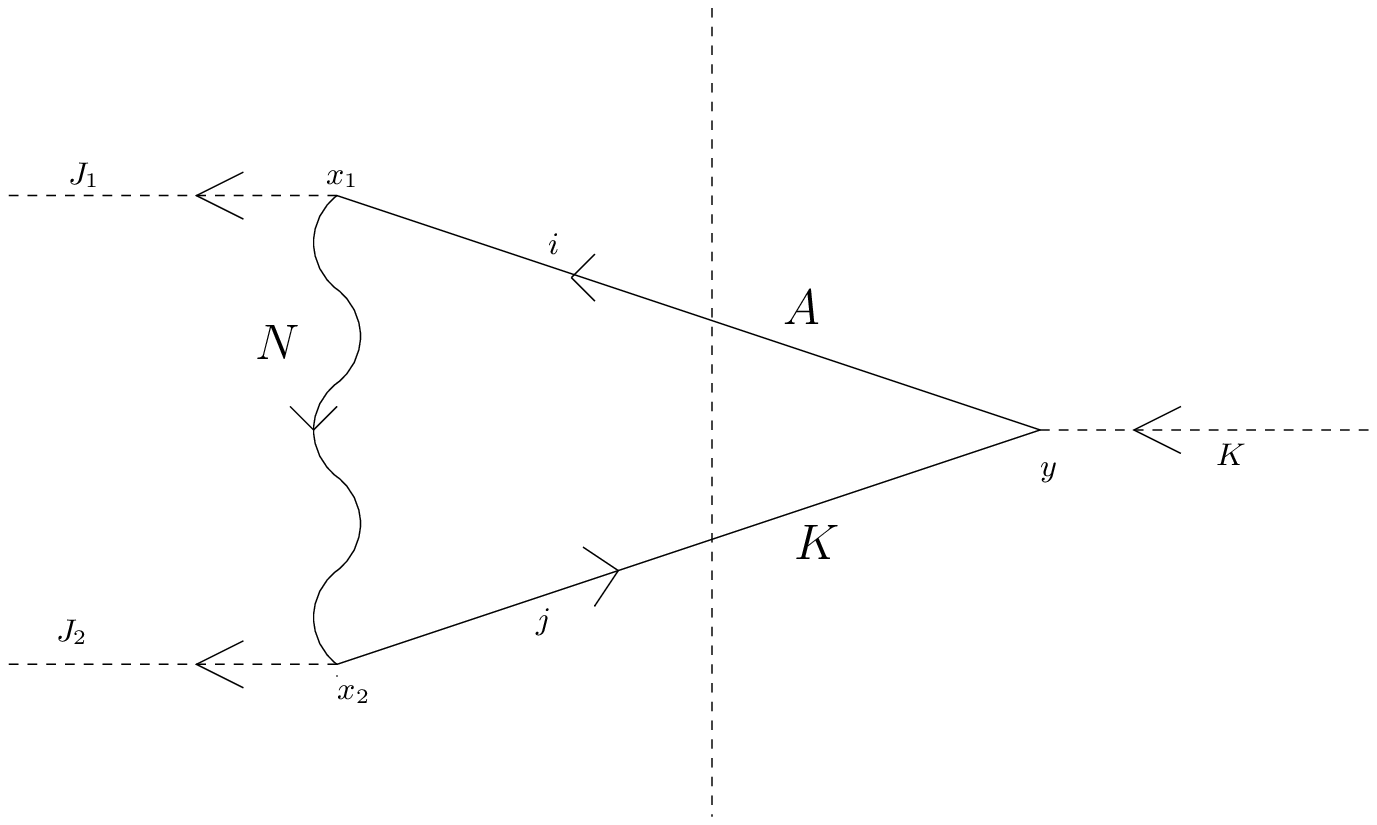}%
\\
fig. 4 : contribution to $W_{2,1}^{c}\left(  x_{1},x_{2};y\right)  $%
\label{fig004}%
\end{center}

In section 3 and in appendix A, we prove that this combination is equal to%
\begin{align}
W_{2,1}^{c}\left(  N,x_{1},x_{2},y\right)   & =\sum_{\left\{  J_{1}%
,\ J_{2},\ K\right\}  =1}^{\infty}\ \frac{1}{x_{1}^{J_{1}+1}}\ \frac{1}%
{x_{2}^{J_{2}+1}}\ \frac{1}{y^{K+1}}\ I_{2}\left(  N,J_{1},J_{2},K\right)
\nonumber\\
& \tag{4.13.a}\\
I_{2}\left(  N,J_{1},J_{2},K\right)   & =-\delta_{J,K}\ \ \left(  1-T_{J_{1}%
}\right)  \left(  1-T_{J_{2}}\right)  \ \frac{1}{J+1}\ \frac{\Gamma\left(
N+J+1\right)  }{\Gamma\left(  N\right)  }\nonumber\\
& \tag{4.13.b}\\
I_{2}\left(  N,J_{1},J_{2},K\right)   & =-\delta_{J,K}\ \frac{4}%
{J+1}\ sh\left(  \frac{J_{1}}{2}\ \frac{\partial}{\partial N}\right)
\ sh\left(  \frac{J_{2}}{2}\ \frac{\partial}{\partial N}\right)  \ N^{\ast
\ast(J+1)}\nonumber\\
& \tag{4.13.c}%
\end{align}
where $J=K=J_{1}+J_{2}.\ $The large $N$ behaviour of

$W_{2,1}^{c}\left(  N,x_{1}\sqrt{N},x_{2}\sqrt{N},y\sqrt{N}\right)  \ $is
found to be%
\begin{align}
& -\frac{y}{N^{\frac{5}{2}}}\sum_{\left\{  J_{1},\ J_{2}\right\}  =1}^{\infty
}\ \frac{1}{\left(  yx_{1}\right)  ^{J_{1}+1}}\ \frac{1}{\left(
yx_{2}\right)  ^{J_{2}+1}}\ \ J_{1}J_{2}J\ast\nonumber\\
& \ast\left[  1+\frac{1}{24N^{2}}\left(  J-1\right)  \left(  J-2\right)
\left\{  \left(  J+1\right)  \left(  J-2\right)  -2J_{1}J_{2}\right\}
+O\left(  \frac{1}{N^{4}}\right)  \right] \nonumber\\
& \tag{4.14}%
\end{align}
or
\begin{equation}
W_{2,1}^{c}\left(  N,x_{1}\sqrt{N},x_{2}\sqrt{N},y\sqrt{N}\right)  =-\frac
{2y}{N^{\frac{5}{2}}}\left[  \frac{y^{2}x_{1}x_{2}-1}{\left(  yx_{1}-1\right)
^{3}\ \left(  yx_{2}-1\right)  ^{3}}+O\left(  \frac{1}{N^{2}}\right)  \right]
\tag{4.15}%
\end{equation}
Finally from (4.13.c), the BMN limit, where \bigskip$N,J_{1},J_{2}$ are large with

$\frac{J_{1}^{2}}{N},\ \frac{J_{2}^{2}}{N}\ $constant, is obtained as%
\begin{equation}
I_{2}\left(  N,J_{1},J_{2},K\right)  \sim-\delta_{J,K}\ \ \frac{4N^{J+1}}%
{J}\ sh\left(  \frac{J_{1}J}{2N}\right)  \ sh\left(  \frac{J_{2}J}{2N}\right)
\tag{4.16}%
\end{equation}

\bigskip

3$%
{{}^\circ}%
)\ $the generalisation of the situation 2$%
{{}^\circ}%
)$ is the calcumlation of

$W_{p,1}^{c}\left(  N,x_{1}...,x_{p},y\right)  $ and is easy to understand. We
sum over $p!$ graphs obtained by exchanging the $p$ points $x_{i}\ $and we
calculate%
\begin{equation}
W_{p,1}^{c}\left(  N,x_{i},y\right)  =\left[  A_{N-1}\left(  y,x_{1}\right)
\
{\displaystyle\prod\limits_{i=1}^{p-1}}
N_{N-1}\left(  x_{i},x_{i+1}\right)  \ K_{N-1}\left(  x_{p},y\right)
+Sym\left(  x_{i}\right)  \right] \tag{4.17}%
\end{equation}
%

\begin{center}
\includegraphics[
height=5.1862cm,
width=7.0951cm
]%
{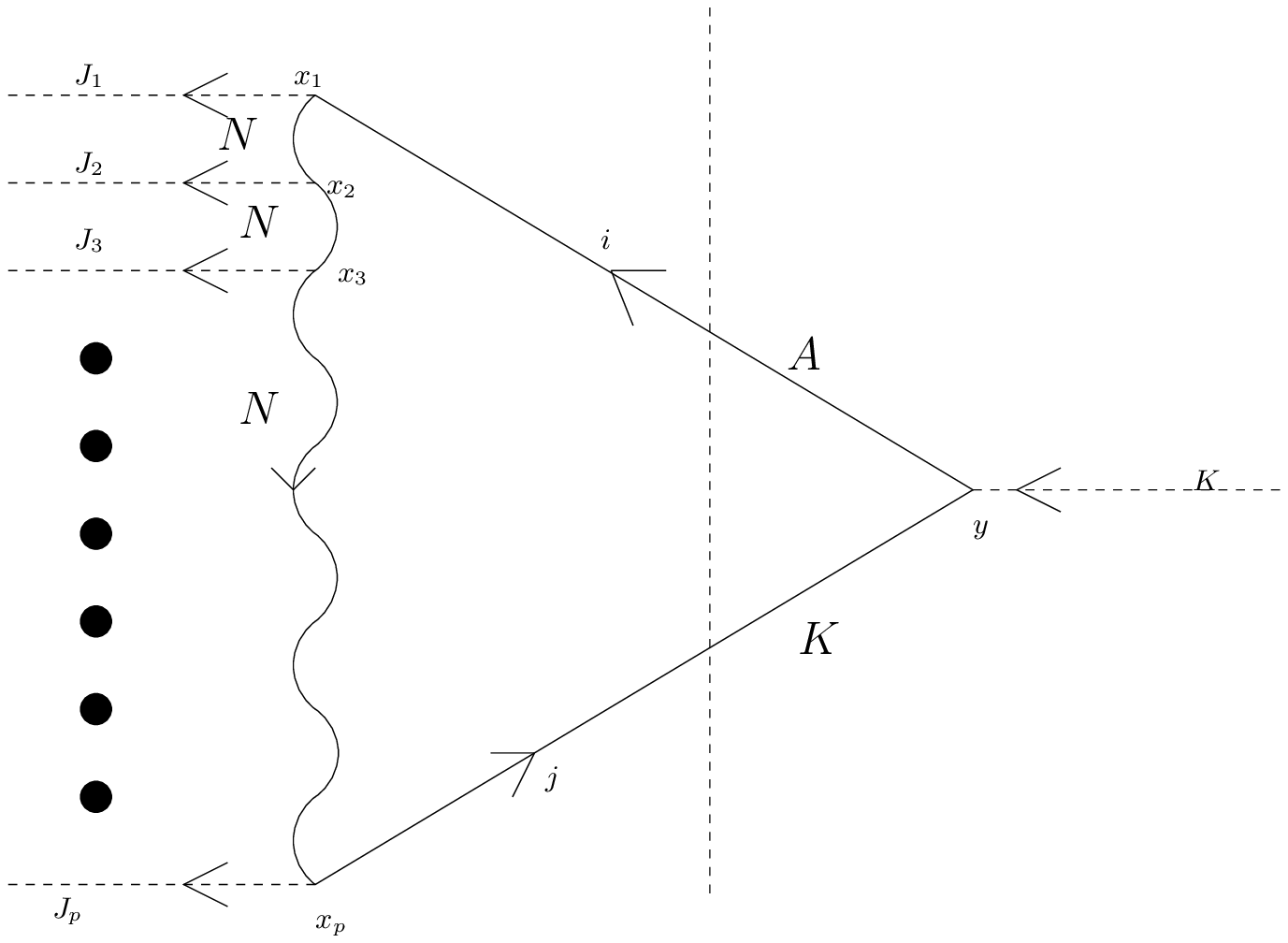}%
\\
fig. 5 : contribution to $W_{p,1}^{c}\left(  x_{1},\ldots,x_{p};y\right)  $%
\label{fig005}%
\end{center}

From section 3 and appendix A, we know that the poles of the propagators
$N_{N-1}\left(  x_{i},x_{i+1}\right)  $ cancel in the sum and that the result
can be written as%
\begin{align}
W_{p,1}^{c}\left(  N,x_{i},y\right)   & =\sum_{\left\{  J_{i},\ K\right\}
=1}^{\infty}%
{\displaystyle\prod\limits_{i=1}^{p}}
\ \frac{1}{x_{i}^{J_{i}+1}}\ \frac{1}{y^{K+1}}\ I_{p}\left(  N,J_{i},K\right)
\tag{4.18.a}\\
I_{p}\left(  N,J_{i},K\right)   & =-\delta_{J,K}\ \
{\displaystyle\prod\limits_{i=1}^{p}}
\left(  1-T_{J_{i}}\right)  \ \frac{1}{J+1}\ \frac{\Gamma\left(  N+J+1\right)
}{\Gamma\left(  N\right)  }\tag{4.18.b}\\
I_{p}\left(  N,J_{i},K\right)   & =-\delta_{J,K}\ \frac{2^{p}}{J+1}\
{\displaystyle\prod\limits_{i=1}^{p}}
sh\left(  \frac{J_{i}}{2}\ \frac{\partial}{\partial N}\right)  \ \ N^{\ast
\ast(J+1)}\tag{4.18.c}%
\end{align}
where $J=K=J_{1}+...+J_{p}.$ The asymptotic behaviour$\ $at large $N$ of

$W\left(  N,x_{i}\sqrt{N},y\sqrt{N}\right)  ,\ $after a proper rescaling of
the variables, is%
\begin{equation}
-\frac{y^{p-1}}{N^{\frac{3p-1}{2}}}\sum_{\left\{  J_{i}\right\}  =1}^{\infty}%
{\displaystyle\prod\limits_{i=1}^{p}}
\ \frac{1}{\left(  x_{i}y\right)  ^{J_{i}+1}}\left[  J_{1}...J_{p}J\left(
J-1\right)  ...\left(  J-p+2\right)  +O\left(  \frac{1}{N^{2}}\right)  \right]
\tag{4.19}%
\end{equation}
or%
\begin{equation}
W_{p,1}^{c}\left(  N,x_{i}\sqrt{N},y\sqrt{N}\right)  =\frac{\left(  -\right)
^{p}}{N^{\frac{3p-1}{2}}}\left\{  \frac{\partial^{p-1}}{\partial y^{p-1}}%
\frac{y^{2p-2}}{%
{\displaystyle\prod\limits_{i=1}^{p}}
\left(  x_{i}y-1\right)  ^{2}\ }+O\left(  \frac{1}{N^{2}}\right)  \right\}
\ \tag{4.20}%
\end{equation}
The BMN limit is trivially obtained as%
\begin{equation}
I_{p}\left(  N,J_{i},K\right)  \sim-\delta_{J,K}\ \frac{2^{p}N^{J+1}}{J}\
{\displaystyle\prod\limits_{i=1}^{p}}
sh\left(  \frac{J_{i}J}{2N}\ \right)  \ \tag{4.21}%
\end{equation}

\bigskip

4$%
{{}^\circ}%
)\ $We now apply the formalism to the four points, connected, resolvant
function $W_{2,2}^{c}\left(  N,x_{1},x_{2},y_{1},y_{2}\right)  $.\ \ First we
calculate the contribution of the graph of fig.6 and its crossed symmetric
ones $\left(  J_{1}\Leftrightarrow J_{2}\ \text{and \ }K_{1}\Leftrightarrow
K_{2}\right)  $.%

\begin{center}
\includegraphics[
height=4.257cm,
width=7.0951cm
]%
{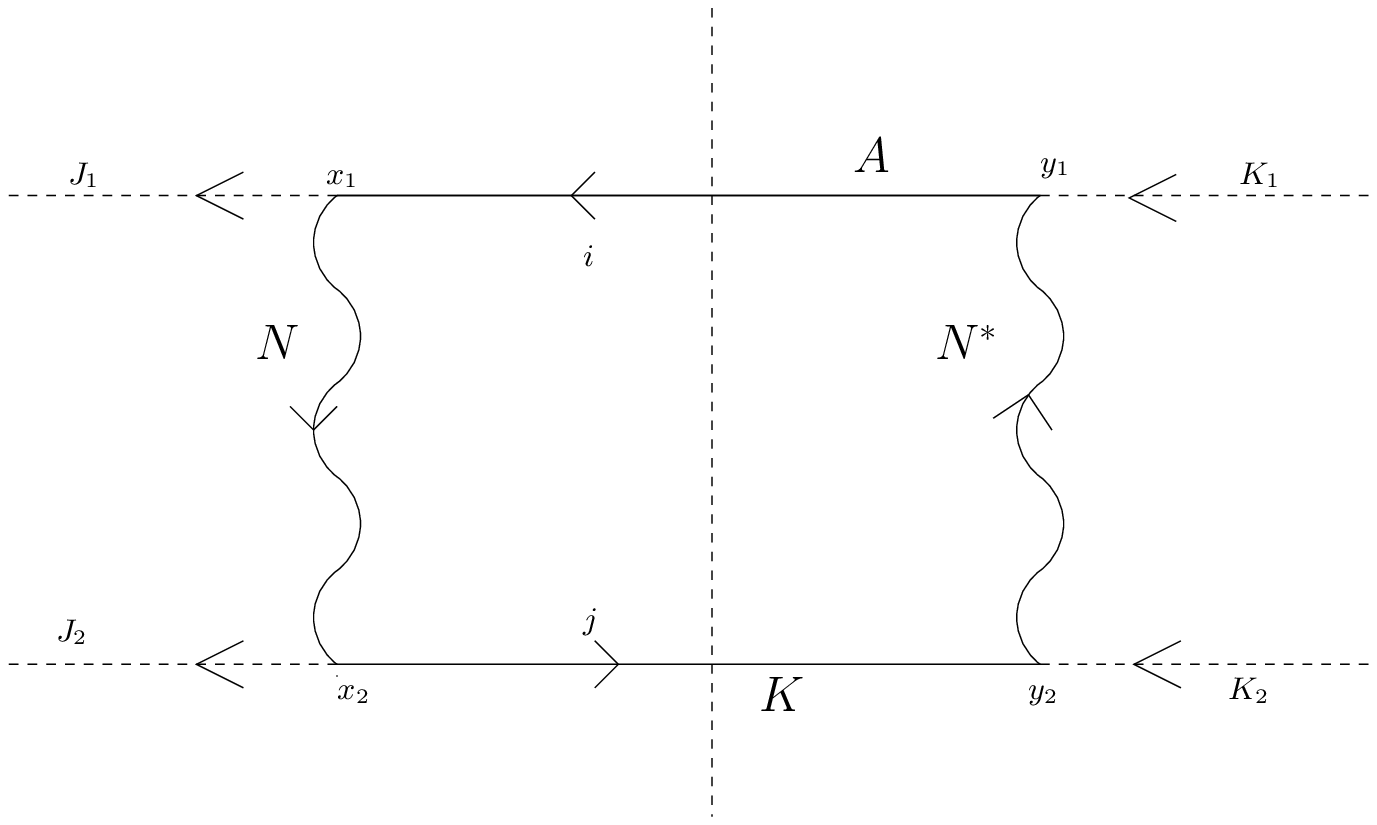}%
\\
fig. 6 : contribution to $W_{2,2}^{c}\left(  x_{1},x_{2};y_{1},y_{2}\right)  $%
\label{fig006}%
\end{center}

We write%
\begin{equation}
W\left(  N,x_{1},x_{2},y_{1},y_{2}\right)  =\sum_{\left\{  J_{i},\ K\right\}
=1}^{\infty}\left(
{\displaystyle\prod\limits_{i=1}^{2}}
\ \frac{1}{x_{i}^{J_{i}+1}}\right)  \ \left(
{\displaystyle\prod\limits_{i=1}^{2}}
\ \frac{1}{y_{i}^{K_{i}+1}}\right)  \ I\left(  N,J_{i},K_{i}\right) \tag{4.22}%
\end{equation}
The formalism of section 3 and appendix A applied to the corresponding four
graphs gives%
\begin{align}
I\left(  N,J_{i},K_{i}\right)   & =-\delta_{J,K}\ O(J_{i},K_{i})\ \ \frac
{1}{J+1}\ \frac{\Gamma\left(  N+J+1\right)  }{\Gamma\left(  N\right)
}\tag{4.23.a}\\
O(J_{i},K_{i})  & =\left[  \left(  1-T_{J_{1}}\right)  \ \left(  1-T_{J_{2}%
}\right)  \otimes\left(  1-T_{K_{1}}\right)  \ \left(  1-T_{K_{2}}\right)
\right] \tag{4.23.b}\\
O(J_{i},K_{i})  & =\left(  1-T_{Inf}\right)  \left(  1+T_{Sup}\right)
\tag{4.23.c}%
\end{align}
where the symbol $\otimes$ is defined in (3.31)\ and where%
\begin{align}
J  & =K=J_{1}+J_{2}=K_{1}+K_{2}\tag{4.24.a}\\
Inf  & =Inf\left(  J_{i},K_{i}\right)  \ \ \ \ \ \ \ \ \ Sup=Sup\left(
J_{i},K_{i}\right) \tag{4.24.b}%
\end{align}

\bigskip The function $I\left(  N,J_{i},K_{i}\right)  $ can also be written%
\begin{equation}
I\left(  N,J_{i},K_{i}\right)  =-\delta_{J,K}\frac{4}{J+1}\ sh\left(
\frac{Inf}{2}\frac{\partial}{\partial N}\right)  \ ch\left(  \frac{Sup}%
{2}\frac{\partial}{\partial N}\right)  \ N^{\ast\ast\left(  J+1\right)
}\tag{4.25}%
\end{equation}
The BMN limit is immediately obtained as%
\begin{equation}
I\left(  N,J_{i},K_{i}\right)  \sim-\delta_{J,K}\frac{4N^{J+1}}{J}\ sh\left(
\frac{J.Inf}{2N}\right)  \ ch\left(  \frac{J.Sup}{2N}\right)  \ \tag{4.26}%
\end{equation}
The large $N$ limit is obtained by expansion of the $sh$ and $ch$ in (4.25) as%
\begin{align}
I\left(  N,J_{i},K_{i}\right)   & =-\delta_{J,K}\ 2\ Inf\ N^{J}\ \ast
\nonumber\\
& \ast\left[  1+\frac{J\left(  J-1\right)  }{24N^{2}}\left[  3\ Sup^{2}%
+Inf^{2}-\left(  J+2\right)  \right]  +O\left(  \frac{1}{N^{4}}\right)
\right] \nonumber\\
& \tag{4.27}%
\end{align}

\bigskip

5$%
{{}^\circ}%
)\ $The remaining part of the connected four points resolvant function
$W_{2,2}^{c}\left(  N,x_{1},x_{2},y_{1},y_{2}\right)  $ is attached to the
skeleton graphs of fig.3 and its crossed symmetric one $\left(  J_{1}%
\Leftrightarrow J_{2}\right)  $ or $\left(  K_{1}\Leftrightarrow K_{2}\right)
$\ .\ According to section 3, we have\ for the graph of fig.3\ \ \
\begin{equation}
I_{1}\left(  N,J_{i},K_{i}\right)  =\delta_{J,K}\ \left(  T_{X}-T_{Y}\right)
\ \sum_{j=0}^{N-1}\ \frac{\Gamma\left(  j+K_{1}+1\right)  }{\Gamma\left(
j+K_{1}-J_{1}+1\right)  }\ \frac{\Gamma\left(  j+J_{2}+1\right)  }%
{\Gamma\left(  j+1\right)  }\tag{4.28}%
\end{equation}
with $J=J_{1}+J_{2}=K_{1}+K_{2}=K$ and with%
\begin{equation}
X=Sup\left(  K_{1}-J_{1},0\right)  \ \ \ \ \ \ \ \ \ \ \ Y=Inf\left(
K_{1},J_{2}\right) \tag{4.29}%
\end{equation}
The sum in (4.28) is transformed in appendix B (B.31) into%
\begin{equation}
\sum_{p=0}^{Inf\left(  J_{1},K_{2}\right)  }\left(  -\right)  ^{p}%
\ \ p!\ \ C_{J_{1}}^{p}\ C_{K_{2}}^{p}\ \frac{\Gamma\left(  N+J-p+1\right)
}{\Gamma\left(  N\right)  }\ \tag{4.30}%
\end{equation}
We can now compute the large $N$ behaviour of $I_{1}\left(  N,J_{i}%
,K_{i}\right)  ;$ for instance, in the sector$\ J_{1}\geq K_{1},K_{2}\geq
J_{2}$ we obtain
\begin{align}
I_{1}\left(  N,J_{i},K_{i}\right)   & =\delta_{J,K}\ J_{2}\ N^{J}\ \left[
1+\frac{J_{1}\left(  K_{1}-K_{2}\right)  }{2N}+\frac{1}{24N^{2}}%
\digamma+O\left(  \frac{1}{N^{3}}\right)  \right] \nonumber\\
& \tag{4.31.a}\\
\digamma & =J\left(  J-1\right)  \left[  3\ J_{1}^{2}+J_{2}^{2}-\left(
J+2\right)  \right]  -12\ K_{1}K_{2}\ J_{1}\left(  J_{1}-1\right) \nonumber\\
& \tag{4.31.b}%
\end{align}
For the BMN\ behaviour, each term in (4.30) contributes, however the sum of
these contributions can be performed and we obtain%
\begin{equation}
I_{1}\left(  N,J_{i},K_{i}\right)  \sim\delta_{J,K}\ \frac{2N^{J+1}}%
{J}\ sh\left(  \frac{J_{2}J}{2N}\right)  \ e^{\frac{J_{1}\left(  K_{1}%
-K_{2}\right)  }{2N}}\tag{4.32}%
\end{equation}

If we sum the graph of fig.3 and the crossed one, and if we describe each of
the sectors as%
\begin{equation}
Sup\geq A,B\geq Inf\tag{4.33}%
\end{equation}
with $A+B=Sup+Inf=J=K,\ \ $we obtain the large $N$ behaviour for both graphs
as%
\begin{align}
I\left(  N,J_{i},K_{i}\right)   & =\delta_{J,K}\ 2\ Inf\ N^{J}\ \left[
1+\frac{1}{24N^{2}}\digamma+O\left(  \frac{1}{N^{4}}\right)  \right]
\tag{4.34.a}\\
\digamma & =J\left(  J-1\right)  \left[  3\ Sup^{2}+Inf^{2}-\left(
J+2\right)  \right]  -12\ AB\ Sup\left(  Sup-1\right) \nonumber\\
& \tag{4.34.b}%
\end{align}
and the BMN behaviour as%
\begin{equation}
I\left(  N,J_{i},K_{i}\right)  \sim\delta_{J,K}\ \frac{4N^{J+1}}{J}\ sh\left(
\frac{J.Inf}{2N}\right)  ch\left(  \frac{\left(  A-B\right)  .Sup}{2N}\right)
\tag{4.35}%
\end{equation}

6$%
{{}^\circ}%
)$ \ Finally, we sum the contributions 4$%
{{}^\circ}%
)$ and 5$%
{{}^\circ}%
)$ to obtain the asymptotic behaviour for the complete connected four points
correlation function; taking into account the relative minus sign between 4$%
{{}^\circ}%
)$ and 5$%
{{}^\circ}%
)$, we obtain%
\begin{align}
\Gamma_{2}^{c}\left(  N,J_{i},K_{i}\right)   & =-\delta_{J,K}\ \ \ N^{J-2}%
\ \ J_{1}J_{2}K_{1}K_{2}\left(  Sup-1\right)  \ +O\left(  N^{J-4}\right)
\nonumber\\
& \tag{4.36.a}\\
\Gamma_{2}^{c}\left(  N,J_{i},K_{i}\right)   & \sim-\delta_{J,K}%
\ \ \frac{8N^{J+1}}{J}\ sh\left(  \frac{J.Inf}{2N}\right)  \ sh\left(
\frac{A.Sup}{2N}\right)  \ sh\left(  \frac{B.Sup}{2N}\right) \nonumber\\
& \tag{4.36.b}%
\end{align}

\bigskip

We close this section with the following remarks.\ Although, in principle, we
should be able to calculate the contribution of any graph, we are unable at
the present time to give the expression for the sum of the graphs which enters
a given $k$-points correlation function; this situation may improve if we use
the so-called loop equations$\left[  12\right]  .\ $From the exemple 5$%
{{}^\circ}%
)$, we see that the t'Hooft topological expansion in $\frac{1}{N^{2}}\ $is not
true graph by graph and that the cancellations between graphs are far from
evident.\ Finally, we verify on these exemples that at large $N$ the resolvant
function properly rescaled behave as%
\begin{equation}
W_{k_{1},k_{2}}\left(  N,x_{i}\sqrt{N},y_{i}\sqrt{N}\right)  \rightarrow
O\left(  N^{\frac{k}{2}}\right) \tag{4.37}%
\end{equation}
because of the disconnected one point self closing loops (the lowest term of
the expansion) and that the connected resolvant function properly rescaled
behave as%
\begin{equation}
W_{k_{1},k_{2}}^{c}\left(  N,x_{i}\sqrt{N},y_{i}\sqrt{N}\right)  \rightarrow
O\left(  \frac{1}{N^{\frac{3k-4}{2}}}\right) \tag{4.38}%
\end{equation}
where $k=k_{1}+k_{2}.$

\appendix

\section{Appendix: summation over the chains of propagators $N$ and $N^{\ast}%
$}

\bigskip

We consider a chain of \ $\left(  p-1\right)  $ propagators $N$ \ \ (the
arguments are the same for the chains of propagators $N^{\ast}$). The
propagator $N_{n}\left(  \xi_{i},\xi_{i+1}\right)  \ $has a pole at $\xi
_{i}=\xi_{i+1}.$ However, when we sum over the $p!$ chains obtained by
exchanging the $p$ points, the poles disappear as we explain now: we consider
two points $\xi_{b}$ and $\xi_{c}$ and we look for the pole at$\ \xi_{b}%
=\xi_{c}$ $;\ $many chains do not contain the propagators $N_{n}\left(
\xi_{b},\xi_{c}\right)  $ or $N_{n}\left(  \xi_{c},\xi_{b}\right)  \ $and have
no pole at $\xi_{b}=\xi_{c}.\ $Now, we associate by pair the chains with a
propagator $N_{n}\left(  \xi_{b},\xi_{c}\right)  $ or $N_{n}\left(  \xi
_{c},\xi_{b}\right)  $ by exchanging $b$ and $c$; the combination$\ \ $

$\left(  N_{n}\left(  \xi_{a},\xi_{b}\right)  \ N_{n}\left(  \xi_{b},\xi
_{c}\right)  \ N_{n}\left(  \xi_{c},\xi_{d}\right)  +N_{n}\left(  \xi_{a}%
,\xi_{c}\right)  \ N_{n}\left(  \xi_{c},\xi_{b}\right)  \ N_{n}\left(  \xi
_{b},\xi_{d}\right)  \right)  \ $has no pole at$\ \xi_{b}=\xi_{c}\ $and is
equal to%
\begin{equation}
...\left(  \frac{\xi_{a}}{\xi_{d}}\right)  ^{n+1}\ \frac{\xi_{d}-\xi_{a}%
}{\left(  \xi_{b}-\xi_{a}\right)  \left(  \xi_{d}-\xi_{c}\right)  \left(
\xi_{c}-\xi_{a}\right)  \left(  \xi_{d}-\xi_{b}\right)  }\ ...\tag{A.1}%
\end{equation}
Similar arguments prove the absence of poles at$\ \xi_{b}=\xi_{c}$ if $b$
or/and $c$ are end-points of the chain, for instance in$\left(  N_{n}\left(
\xi_{b},\xi_{c}\right)  \ +N_{n}\left(  \xi_{c},\xi_{b}\right)  \right)  $ or
in\newline$\left(  N_{n}\left(  \xi_{a},\xi_{b}\right)  \ N_{n}\left(  \xi
_{b},\xi_{c}\right)  \ +N_{n}\left(  \xi_{a},\xi_{c}\right)  \ N_{n}\left(
\xi_{c},\xi_{b}\right)  \right)  .$ We remind that the self closing loops
$N_{n}\left(  \xi_{b},\xi_{b}\right)  \ $are replaced in (1.26) by
$H_{n}\left(  \xi_{b},\xi_{b}\right)  \ $given in (3.6.a) and contribute only
to disconnected parts of the graphs $(J_{i}=0)$ and that the two points
loop$\ N_{n}\left(  \xi_{b},\xi_{c}\right)  N_{n}\left(  \xi_{c},\xi
_{b}\right)  \ $is replaced in (1.26) by $"N_{n}\left(  \xi_{b},\xi
_{c}\right)  N_{n}\left(  \xi_{c},\xi_{b}\right)  "$ which has no pole and is
zero in rotationnally invariant systems.

\bigskip

Consequently, since we have no poles at coinciding $\xi^{\prime}s$, the
expansion in $\frac{1}{\xi_{i}}$ $\ $for the contribution of the $p!$ chains
is the same whatever sector we choose $\xi_{\alpha_{1}}\leq\xi_{\alpha_{2}%
}\leq...\leq\xi_{\alpha_{p}}\ $and the results must contain terms of the type
$\left(  \frac{1}{\xi_{i}}\right)  ^{J_{i}+1}$ with $J_{i}>0\ $if $p>1.$
According to the chosen sector we use the convergent expansion%
\begin{align}
\frac{1}{\xi_{i+1}-\xi_{i}}  & =\frac{1}{\xi_{i+1}}\sum_{j=0}^{\infty}\left(
\frac{\xi_{i}}{\xi_{i+1}}\right)  ^{j}\ ,\ \ \ \ \ \xi_{i}<\ \xi
_{i+1}\tag{A.2.a}\\
\frac{1}{\xi_{i+1}-\xi_{i}}  & =-\frac{1}{\xi_{i}}\sum_{j=0}^{\infty}\left(
\frac{\xi_{i+1}}{\xi_{i}}\right)  ^{j}\ ,\ \ \ \ \ \ \xi_{i}>\xi
_{i+1}\tag{A.2.b}%
\end{align}
Now, each chain separetely can be expanded in the choosen sector but if we
find in the expansion some contributions $\left(  \frac{1}{\xi_{i}}\right)
^{J_{i}+1}$ with a negative or nul power $J_{i}$, we may ignore that
contribution since we know that it must be cancelled when we sum over the $p!
$ chains.

For instance, we consider three consecutive points of the chain and the
corresponding situation%
\begin{equation}
N_{n}\left(  \xi_{i-1},\xi_{i}\right)  \ N_{n}\left(  \xi_{i},\xi
_{i+1}\right)  ,\ \ \ \ \ \ \ \xi_{i}=Inf\left(  \xi_{i-1},\xi_{i},\xi
_{i+1}\right) \tag{A.3}%
\end{equation}
according to (A.2) \ we obtain the contribution%
\begin{equation}
-\frac{1}{\xi_{i-1}\ \xi_{i+1}}\ \left(  \frac{\xi_{i-1}}{\xi_{i+1}}\right)
^{n+1}\ \sum_{r,s=0}^{\infty}\ \left(  \frac{\xi_{i}}{\xi_{i-1}}\right)
^{r}\ \left(  \frac{\xi_{i}}{\xi_{i+1}}\right)  ^{s}\tag{A.4}%
\end{equation}
This term provides for the variable $\xi_{i}$ a contribution
\begin{equation}
\left(  \frac{1}{\xi_{i}}\right)  ^{-r-s}\rightarrow J_{i}=-r-s-1<0\tag{A.5}%
\end{equation}
so that this chain should be ignored in the choosen sector. In such a sector,
only the chains with no other relative minimum that the end-points of the
chain should be considered.\ \ 

A first important consequence of this remark is that the chains which form a
loop gives no contribution; this can be verified for instance in (A.1) with
$a=d$\bigskip\ in the case of a three points loop. A second consequence is
that if we compute%
\begin{equation}
-A_{n}\left(  \eta_{1}^{\ast},\xi_{b_{1}}\right)
{\displaystyle\prod\limits_{i=1}^{p-1}}
N_{n}\left(  \xi_{b_{i}},\xi_{b_{i+1}}\right)  \ K_{n}\left(  \xi_{b_{p}}%
,\eta_{2}^{\ast}\right) \tag{A.6}%
\end{equation}
in the sector $\xi_{\alpha_{1}}\leq\xi_{\alpha_{2}}\leq...\leq\xi_{\alpha_{p}%
}\ \ $then$\ ,\ \xi_{\alpha_{1}}$ is either $\xi_{b_{1}}$ or $\xi_{b_{p}}$ and
along the chain we have
\begin{equation}
\xi_{b_{1}}\leq\xi_{b_{2}}\leq...\leq\xi_{b_{q}}=\xi_{\alpha_{p}}\geq
\xi_{b_{q+1}}\geq\xi_{b_{q+2}}\geq...\geq\xi_{b_{p}}\ \ \ \ \ \ \ 1\leq q\leq
p\tag{A.7}%
\end{equation}
From (A.2), we obtain for (A.6)%
\begin{align}
& \left(  -\right)  ^{p-q}\frac{1}{\left(  \eta_{1}^{\ast}\right)  ^{i+1}%
}\ \left(  \frac{1}{\xi_{b_{1}}}\right)  ^{i-n-r_{1}}%
{\displaystyle\prod\limits_{i=2}^{q-1}}
\left(  \frac{1}{\xi_{b_{i}}}\right)  ^{r_{i-1}-r_{i}+1}\ast\nonumber\\
& \ast\left(  \frac{1}{\xi_{b_{q}}}\right)  ^{r_{q-1}+r_{q}+2}%
{\displaystyle\prod\limits_{i=q+1}^{p-1}}
\left(  \frac{1}{\xi_{b_{i}}}\right)  ^{r_{i}-r_{i-1}+1}\left(  \frac{1}%
{\xi_{b_{p}}}\right)  ^{n-j-r_{p-1}+1}\frac{1}{\left(  \eta_{2}^{\ast}\right)
^{-j}}\nonumber\\
& \tag{A.8}%
\end{align}
and by identification, we have%
\begin{align}
J_{b_{1}}  & =i-n-r_{1}-1\tag{A.9.a}\\
J_{b_{i}}  & =r_{i-1}-r_{i}\ \ \ \ \ \ \ \ \ \ \ \ \ \ 2\leq i\leq
q-1\tag{A.9.b}\\
J_{b_{q}}  & =r_{q-1}+r_{q}+1\tag{A.9.c}\\
J_{b_{i}}  & =r_{i}-r_{i-1}\ \ \ \ \ \ \ \ \ \ \ \ q+1\leq i\leq
p-1\tag{A.9.d}\\
J_{b_{p}}  & =n-j-r_{p-1}\tag{A.9.e}%
\end{align}
which implies%
\begin{equation}
J=\sum_{i=1}^{p}J_{b_{i}}=i-j\tag{A.10}%
\end{equation}
Clearly, the variables $i$ and $r_{i}$ are all determined as function
of$\ n,j$ and\ $J_{b_{i}};$ the constraints $i\geq n+1$ and $r_{i}\geq0$
together with $J_{b_{i}}>0\ $implies only two constraints on $j$%
\begin{align}
j  & \leq n-J_{+}\tag{A.11.a}\\
j  & \geq n-J_{+}-J_{b_{q}}+1\tag{A.11.b}%
\end{align}
with%
\begin{equation}
J_{+}=\sum_{i=q+1}^{p}J_{b_{i}}\tag{A.12}%
\end{equation}
The original sum over $j$ of the function $\frac{h_{i}}{h_{j}}\ $describing
the kernels $A_{n}\left(  \eta_{1}^{\ast},\xi_{b_{1}}\right)  \ $and
$K_{n}\left(  \xi_{b_{p}},\eta_{2}^{\ast}\right)  $ is, for that chain,
transformed into%
\begin{equation}
\sum_{j=0}^{n}\frac{h_{j+J}}{h_{j}}...\Rightarrow\left(  -\right)  ^{p-q}%
\sum_{j=n-J_{+}-J_{b_{q}}+1}^{n-J_{+}}\frac{h_{j+J}}{h_{j}}\ =\left(
-\right)  ^{p-q}\ T_{J_{+}}\left(  1-T_{J_{a_{p}}}\right)  \sum_{j=0}^{n}%
\frac{h_{j+J}}{h_{j}}...\tag{A.13}%
\end{equation}
where the operator $T_{a}$ is defined in (3.16).

\ \bigskip Finally, summing over all the $p!$ chains consists in summing over
all subsets of $\left\{  \alpha_{1},...,\alpha_{p-1}\right\}  $(including the
empty set and the complete set) corresponding to the definition of $J_{+}%
.\ $Taking into account the alternate sign in $\left(  -\right)  ^{p-q}$ the
total sum is nothing but%
\begin{equation}%
{\displaystyle\prod\limits_{i=1}^{p}}
\left(  1-T_{J_{i}}\right)  \ \sum_{j=0}^{n}\frac{h_{j+J}}{h_{j}}...\tag{A.14}%
\end{equation}
This result is the purpose of this appendix.

\section{Appendix: the 't Hooft expansion}

\bigskip

We define the symbol%
\begin{equation}
n^{\ast J}=\left(  n+1\right)  \left(  n+2\right)  ...\left(  n+J\right)
\tag{B.1}%
\end{equation}
which is expanded in powers of $n$ as%
\begin{equation}
n^{\ast J}=\sum_{p=0}^{J}\ \sigma_{p}\left(  J\right)  \ n^{J-p}\tag{B.2}%
\end{equation}
with%
\begin{align}
\sigma_{0}\left(  J\right)   &
=1\ \ \ \ \ \ \ \ \ \ \ \ \ \ \ \ \ \ \ \ \ \ \ \ \ \ \sigma_{J}\left(
J\right)  =J!\tag{B.3.a}\\
\sigma_{1}\left(  J\right)   & =\frac{J\left(  J+1\right)  }{2}%
\ \ \ \ \ \ \ \ \ \ \ \ \ \sigma_{p}\left(  J\right)  =0\ \ \ \ \text{if\ }%
p>J\tag{B.3.b}%
\end{align}
For a given $p,\ $the coefficients $\sigma_{p}\left(  J\right)  $ are
polynomials in $J;\ $clearly we have%
\begin{equation}
\sigma_{p}\left(  J\right)  =\sum_{1\leq k_{1}<k_{2}<...<k_{p}\leq J}%
\ k_{1}\ k_{2}...k_{p}\tag{B.4}%
\end{equation}
and we observe the relation%
\begin{equation}
\sigma_{p}\left(  J\right)  =\sigma_{p}\left(  J-1\right)  +J\ \sigma
_{p-1}\left(  J-1\right) \tag{B.5}%
\end{equation}
Since $\sigma_{p}\left(  J\right)  $ vanishes if $\ J<p,$ we write%
\begin{align}
\sigma_{p}\left(  J\right)   & =\frac{1}{2^{p}\ p!}J\left(  J-1\right)
...\left(  J-p+1\right)  \ \pi_{p}\left(  J\right) \tag{B.6.a}\\
\pi_{0}\left(  J\right)   & =1\ \ \ \ \ \ \ \ \ \ \ \ \ \ \ \ \ \ \ \ \ \pi
_{1}\left(  J\right)  =J+1\tag{B.6.b}%
\end{align}
The relation%
\begin{equation}
J\ \pi_{p}\left(  J\right)  =\left(  J-p\right)  \ \pi_{p}\left(  J-1\right)
\ +2pJ\ \pi_{p-1}\left(  J-1\right) \tag{B.7}%
\end{equation}
shows that $\pi_{p}\left(  J\right)  $ is a monic polynomial of degree $p$ in
$J$ which vanishes at $J=-1\ $if $p>0.$We find%
\begin{align}
\pi_{2}\left(  J\right)   & =\left(  J+1\right)  \left(  J+\frac{2}{3}\right)
\ \ \ \ \ \ \ \ \ \ \pi_{3}\left(  J\right)  =J\left(  J+1\right)
^{2}\tag{B.8.a}\\
\pi_{4}\left(  J\right)   & =\left(  J+1\right)  \left(  J^{3}+J^{2}-\frac
{2}{3}J-\frac{8}{15}\right)  \ \tag{B.8.b}\\
\pi_{5}\left(  J\right)   & =J\left(  J+1\right)  ^{2}\left(  J^{2}-\frac
{1}{3}J-2\right)  \ \ \ \ \ \tag{B.8.c}%
\end{align}
We note that at large $J\ $%
\begin{equation}
\sigma_{p}\left(  J\right)  \sim\frac{1}{p!}\ \left(  \frac{J^{2}}{2}\right)
^{p}\tag{B.9}%
\end{equation}
so that the so called "BMN" limit $(n\ $and\ $J\rightarrow\infty,\ \frac
{J^{2}}{n}=cte)\ $is%
\begin{equation}
n^{\ast J}\sim n^{J}\ e^{\frac{J^{2}}{2n}}\tag{B.10}%
\end{equation}

\bigskip

From the generating functional%
\begin{equation}
\left(  1+x\right)  ^{-j-1}=\sum_{J=0}^{\infty}\left(  -\right)  ^{J}\ j^{\ast
J}\ \frac{x^{J}}{J!}\tag{B.11}%
\end{equation}
and from%
\begin{equation}
\sum_{j=0}^{n}\left(  1+x\right)  ^{-j-1}=\frac{1-\left(  1+x\right)  ^{-n-1}%
}{x}\tag{B.12}%
\end{equation}
we obtain by identification of the powers of $x$%
\begin{equation}
\sum_{j=0}^{n}\ j^{\ast J}=\frac{n^{\ast\left(  J+1\right)  }}{J+1}\tag{B.13}%
\end{equation}

\bigskip

In section 4 we have to calculate%
\begin{equation}
\sum_{j=N-J}^{N-1}\ j^{\ast J}=\frac{\left(  N-1\right)  ^{\ast\left(
J+1\right)  }-\left(  N-J-1\right)  ^{\ast\left(  J+1\right)  }}%
{J+1}\tag{B.14}%
\end{equation}
From%
\begin{align}
\left(  N-1\right)  ^{\ast\left(  J+1\right)  }  & =N\left(  N+1\right)
...\left(  N+J\right)  =\sum_{p=0}^{J}\sigma_{p}\left(  J\right)
\ N^{J-p+1}\nonumber\\
& \tag{B.15.a}\\
\left(  N-J-1\right)  ^{\ast\left(  J+1\right)  }  & =N\left(  N-1\right)
...\left(  N-J\right)  =\sum_{p=0}^{J}\left(  -\right)  ^{p}\sigma_{p}\left(
J\right)  \ N^{J-p+1}\nonumber\\
& \tag{B.15.b}%
\end{align}
we obtain the topological t'Hooft expansion\
\begin{equation}
\sum_{j=N-J}^{N-1}\ j^{\ast J}=\frac{2}{J+1}\sum_{p\ odd}^{J}\sigma_{p}\left(
J\right)  \ N^{J-p+1}\tag{B.16}%
\end{equation}
Also, from (B.9), the BMN limit (large $N$, large $J$, $\frac{J^{2}}{N}=cte$)
is%
\begin{equation}
\sum_{j=N-J}^{N-1}\ j^{\ast J}\sim\frac{2N^{J+1}}{J}\ sh\left(  \frac{J^{2}%
}{2N}\right) \tag{B.17}%
\end{equation}

\bigskip

We find convenient to introduce the expression%

\begin{align}
n^{\ast\ast\left(  J+1\right)  }  & =e^{-\left(  \frac{J}{2}+1\right)
\frac{\partial}{\partial n}}\ \ \ n^{\ast\left(  J+1\right)  }\tag{B.18.a}\\
n^{\ast\ast\left(  J+1\right)  }  & =\left(  n-\frac{J}{2}\right)  \left(
n-\frac{J}{2}+1\right)  ...\left(  n+\frac{J}{2}-1\right)  \left(  n+\frac
{J}{2}\right) \nonumber\\
& \tag{B.18.b}%
\end{align}
which has clearly a $\frac{1}{n^{2}}\ $expansion\bigskip%

\begin{equation}
n^{\ast\ast\left(  J+1\right)  }=\sum_{p=0}^{E\left(  \frac{J+1}{2}\right)
}\ \left(  -\right)  ^{p}\ \Sigma_{2p}\left(  J\right)  \ \ n^{J-2p+1}%
\tag{B.19}%
\end{equation}
with%
\begin{equation}
\Sigma_{0}\left(  J\right)  =1\ \ \ \ \ \ \ \ \ \ \ \ \Sigma_{2}\left(
J\right)  =\frac{J\left(  J+1\right)  \left(  J+2\right)  }{24}\tag{B.20}%
\end{equation}
We have the relation%
\begin{equation}
\Sigma_{2p}\left(  J\right)  =\Sigma_{2p}\left(  J-2\right)  +\frac{J^{2}}%
{4}\ \Sigma_{2p-2}\left(  J-2\right) \tag{B.21}%
\end{equation}
We write for $p>0$%
\begin{align}
\Sigma_{2p}\left(  J\right)   & =\frac{\left(  J-2p+2\right)  \left(
J-2p+3\right)  ...J\left(  J+1\right)  }{\left(  24\right)  ^{p}%
\ \ p!}\ \ \theta_{2p}\left(  J\right)  \ \ \ \tag{B.22.a}\\
\theta_{0}\left(  J\right)   &
=1\ \ \ \ \ \ \ \ \ \ \ \ \ \ \ \ \ \ \ \ \ \ \ \ \ \ \theta_{2}\left(
J\right)  =J+2\tag{B.22.b}%
\end{align}
the relation (B.21) becomes%
\begin{equation}
J\left(  J+1\right)  \ \ \theta_{2p}\left(  J\right)  =\left(  J-2p\right)
\left(  J-2p+1\right)  \ \theta_{2p}\left(  J-2\right)  +6pJ^{2}%
\ \theta_{2p-2}\left(  J-2\right) \tag{B.23}%
\end{equation}
The polynomials $\theta_{2p}\left(  J\right)  $ are monic polynomials of
degree $p$ in $J$ which vanish for $J=-2$ if $p>0.\ $We have%
\begin{align}
\theta_{4}\left(  J\right)   & =\left(  J+2\right)  \left(  J+\frac{12}%
{5}\right) \tag{B.24.a}\\
\theta_{6}\left(  J\right)   & =\left(  J+2\right)  \left(  J^{2}+\frac{26}%
{5}J+\frac{48}{7}\right) \tag{B.24.b}\\
\theta_{8}\left(  J\right)   & =\left(  J+2\right)  \left(  J^{3}+\frac{42}%
{5}J^{2}+\frac{4184}{175}J+\frac{576}{25}\right)  \ \ \ \ \tag{B.24.c}%
\end{align}

At large $J$%
\begin{equation}
\Sigma_{2p}\left(  J\right)  \sim\frac{1}{p!}\left(  \frac{J^{3}}{24}\right)
^{p}\tag{B.25}%
\end{equation}
so that the "BMN" limit (and the first correction) of $n^{\ast\ast\left(
J+1\right)  }$ is%
\begin{equation}
n^{\ast\ast\left(  J+1\right)  }\sim n^{J+1\ }\ \ e^{-\frac{J^{3}}{24\ n^{2}}%
}=n^{J+1\ }\left[  1+O\left(  \frac{1}{J}\right)  \right]
,\ \ \ \ \ \ \ \ \ \ \ \ \ \frac{J^{2}}{n}=cte\tag{B.26}%
\end{equation}
the exponential term being the first correction to the "BMN" limit.

\bigskip

With these notations, we may compute directly the $\frac{1}{N^{2}}\ $expansion
of%
\begin{equation}%
{\displaystyle\prod\limits_{i=1}^{p}}
\left(  1-T_{J_{i}}\right)  \ \ \sum_{k=0}^{N-1}\ \frac{\Gamma\left(
k+J+1\right)  }{\Gamma\left(  k+1\right)  }=\frac{2^{p}}{J+1}%
{\displaystyle\prod\limits_{i=1}^{p}}
\left[  sh\left(  \frac{J_{i}}{2}\frac{\partial}{\partial N}\right)  \right]
\ N^{\ast\ast\left(  J+1\right)  }\ \ \ \ \ \ \ \ \ \ \ \ \ \ \ \ \ \tag{B.27}%
\end{equation}
where $\ \ \sum_{i=1}^{p}J_{i}=J;$ the "BMN" limit is trivially%
\begin{equation}%
{\displaystyle\prod\limits_{i=1}^{p}}
\left(  1-T_{J_{i}}\right)  \ \ \sum_{k=0}^{N-1}\ \frac{\Gamma\left(
k+J+1\right)  }{\Gamma\left(  k+1\right)  }\sim2^{p}\ \frac{N^{J+1}}{J}%
{\displaystyle\prod\limits_{i=1}^{p}}
\left[  sh\left(  \frac{J_{i}J}{2N}\right)  \right]  \ \ ,\ \ \ \frac{J_{i}%
J}{N}=cte\tag{B.28}%
\end{equation}

\bigskip

Finally, if we compare the coefficients of the term $x^{b}$ in the expansions%
\begin{equation}
\left(  1+x\right)  ^{-j-1}=\left(  1+x\right)  ^{-j-a-1}\ \left(  1+x\right)
^{a}\tag{B.29}%
\end{equation}
we obtain the relation%
\begin{equation}
\frac{\Gamma\left(  j+b+1\right)  }{\Gamma\left(  j+1\right)  }=\sum
_{q=0}^{Inf\left(  a,b\right)  }\ \left(  -\right)  ^{q}\ q!\ C_{a}^{q}%
\ C_{b}^{q}\ \frac{\Gamma\left(  j+a+b-q+1\right)  }{\Gamma\left(
j+a+1\right)  }\tag{B.30}%
\end{equation}
if $a$ is a non negative integer (with trivial analytic continuation to any
$a$). As a consequence%
\begin{align}
& \sum_{j=0}^{N-1}\frac{\Gamma\left(  j+a+c+1\right)  }{\Gamma\left(
j+1\right)  }\frac{\Gamma\left(  j+b+c+1\right)  }{\Gamma\left(  j+c+1\right)
}\nonumber\\
& =\sum_{q=0}^{Inf\left(  a,b\right)  }\ \left(  -\right)  ^{q}\frac
{q!\ C_{a}^{q}\ C_{b}^{q}}{\left(  a+b+c-q+1\right)  }\ \ \frac{\Gamma\left(
N+a+b+c-q+1\right)  }{\ \Gamma\left(  N\right)  }\nonumber\\
& \tag{B.31}%
\end{align}

\bigskip

\bigskip

\bigskip

\textit{Acknowledgements:\ }I wish to thank B.\ Eynard and C.F.\ Kristjansen
who suggested this calculation.

\end{document}